\documentclass[12pt]{article}
\usepackage{amsfonts}
\usepackage{amssymb}
\usepackage{amsthm}
\usepackage[fleqn]{amsmath}
\usepackage[mathcal]{euscript}
\usepackage{mathrsfs}
\newcommand{\Ref}[1]{(\ref{#1})}

\makeatletter
\renewcommand \thesection {\@arabic\c@section}
\renewcommand\thesubsection   {\thesection.\@arabic\c@subsection}
\renewcommand\thesubsubsection{\thesubsection .\@arabic\c@subsubsection}
\renewcommand\theparagraph    {\thesubsubsection.\@arabic\c@paragraph}
\renewcommand\section{\@startsection {section}{1}{\z@}%
                                   {-3.5ex \@plus -1ex \@minus -.2ex}%
                                   {1.9ex \@plus.2ex}%
                                   {\normalfont\large\bfseries\centering}}
\renewcommand\subsection{\@startsection{subsection}{2}{\z@}%
                                     {-2ex\@plus -1ex \@minus -.2ex}%
                                     {1.2ex \@plus .2ex}%
                                    {\normalfont\normalsize\bfseries\centering}}
\renewcommand\subsubsection{\@startsection{subsubsection}{3}{\z@}%
                                     {-2ex\@plus -1ex \@minus -.2ex}%
                                     {.5ex \@plus .2ex}%
                                     {\normalfont\normalsize\em}}
\renewcommand\paragraph{\@startsection{paragraph}{4}{\z@}%
                                    {3.25ex \@plus1ex \@minus.2ex}%
                                    {-1em}%
                                    {\normalfont\normalsize\em}}
\renewcommand\subparagraph{\@startsection{subparagraph}{5}{\parindent}%
                                       {3.25ex \@plus1ex \@minus .2ex}%
                                       {-1em}%
                                      {\normalfont\normalsize\em}}
\makeatother

\newcommand\abs[1]{{\left| #1 \right|}}
\newcommand\norm[1]{{\left\| #1 \right\|}}

\newcounter{subequation}
	\newenvironment{subequation}%
	{\addtocounter{equation}{-1}%
	\stepcounter{subequation}%
	\begin{equation}}%
	{\end{equation}%
}

\newcommand{\beq}{\begin{equation}}
\newcommand{\eeq}{\end{equation}}
\newcommand{\bseq}{\begin{subequation}}
\newcommand{\eseq}{\end{subequation}}
\newcommand{\bea}{\begin{eqnarray}}
\newcommand{\eea}{\end{eqnarray}}

\newcommand{\sign}{{\mathrm{sign\, }}}

\newcommand{\pmk}{{{z_k}}}

\newcommand{\eps}{\epsilon}

\newcommand{\pr}{\prime}

\newcommand{\QED}{$\quad$\textrm{Q.E.D.}}

\newcommand{\dd}{{\mathrm{d}}}

\newcommand{\cE}{{\cal E}}
\newcommand{\cF}{{\cal F}}
\newcommand{\cG}{{\cal G}}
\newcommand{\cK}{{\cal K}}
\newcommand{\cT}{{\cal T}}

\newcommand{\Lset}{{\mathbb L}}
\newcommand{\Mset}{{\mathbb M}}
\newcommand{\Nset}{{\mathbb N}}
\newcommand{\Rset}{{\mathbb R}}
\newcommand{\Zset}{{\mathbb Z}}

\newcommand{\admA}{{\mathscr{A}}}

\newcommand{\MTWtens}[1]{{\textbf{\textsf{#1}}}}

\newcommand{\dQ}{{\MTWtens{d}}}
\newcommand{\gQ}{{\MTWtens{g}}}

\newcommand{\SPvec}[1]{{\textbf{\textsl{#1}}}}

\newcommand{\xV}{{\SPvec{x}}}

\newcommand{\Tay}{{\mathrm{Tay}}}

\newcommand{\Omc}{\Omega_{\mathrm{crit}}}

\newcommand{\Lloc}{L_{\mathrm{loc}}}

\newtheorem{defn}{Definition}[section]

\newtheorem{Coro}[defn]{Corollary}

\newtheorem{Prop}[defn]{Proposition}
\newtheorem{Rema}[defn]{Remark}
\newtheorem{Theo}[defn]{Theorem}

\begin{document}

\title{ON THE QUASI-LINEAR ELLIPTIC PDE\\
      $-\nabla\cdot(\nabla{u}/\sqrt{1-|\nabla{u}|^2}) = 4\pi\sum_k a_k \delta_{s_k}$\\
	IN PHYSICS AND GEOMETRY}
\author{\sc{ Michael K.-H. Kiessling}\\
	Department of Mathematics\\
	Rutgers, The State University of New Jersey\\
	110 Frelinghuysen Rd., Piscataway, NJ 08854}
\date{Revised version of November 24, 2011}
\maketitle

\begin{abstract}
\noindent
	It is shown that for each finite number $N$ of Dirac measures $\delta_{s_n}$ supported 
at points $s_n\in\Rset^3$ with given amplitudes $a_n\in\Rset\backslash\{0\}$ there exists a unique 
real-valued function $u\in C^{0,1}(\Rset^3)$, vanishing at infinity, which distributionally
solves the quasi-linear elliptic partial differential equation of divergence form
$-\nabla\cdot(\nabla{u}/\sqrt{1-|\nabla{u}|^2}) = 4\pi\sum_{n=1}^N a_n \delta_{s_n}$.
	Moreover, $u\in C^{\omega}(\Rset^3\backslash\{s_n\}_{n=1}^N)$.
The result can be interpreted in at least two ways:
(a) for any number $N$ of point charges of arbitrary magnitude and sign at prescribed locations $s_n$ 
in three-dimensional Euclidean space there exists a unique electrostatic field which satisfies 
the Maxwell-Born-Infeld field equations smoothly away from the point charges and vanishes  as $|s|\to\infty$;
(b) for any number $N$ of integral mean curvatures assigned to locations $s_n\in \Rset^3\subset\Rset^{1,3}$ 
there exists a unique asymptotically flat, almost everywhere space-like maximal slice with point defects of 
Minkowski spacetime $\Rset^{1,3}$, having lightcone singularities over the $s_n$ but being smooth otherwise, 
and whose height function vanishes as $|s|\to\infty$.
	No struts between the point singularities ever occur.
\end{abstract} 

\noindent
{\textbf{Keywords:}}

\noindent
\textit{Geometry}: Lorentz manifolds, maximal foliations, lightcone singularities;

\noindent
\textit{Relativity}: Special relativity, Minkowski spacetime, Poincar\'e transformations;

\noindent
\textit{Electromagnetism}: Maxwell-Born-Infeld equations, point charges, electrostatic fields. 
\smallskip

\hrule
\smallskip
\noindent
\copyright{2011} The author.  
Reproduction of this paper, in its entirety and for noncommercial purposes only, is permitted.

\newpage
%%%%%%%%%%%%%%%%%%%%%%%%%%%%%%%%%%%%%%%%%%%%%%%%%%%%%%%%%%%%%%%%%%%%
%%%%%%%%%%%%%%%%%%%%%%%%%%%%%%%%%%%%%%%%%%%%%%%%%%%%%%%%%%%%%%%%%%%%
%%%%%%%%%%%%%%%%%%%%%%%%%%%%%%%%%%%%%%%%%%%%%%%%%%%%%%%%%%%%%%%%%%%%
%%%%%%%%%%%%%%%%%%%%%%%%%%%%%%%%%%%%%%%%%%%%%%%%%%%%%%%%%%%%%%%%%%%%
	\section{Introduction}
%%%%%%%%%%%%%%%%%%%%%%%%%%%%%%%%%%%%%%%%%%%%%%%%%%%%%%%%%%%%%%%%%%%%
%%%%%%%%%%%%%%%%%%%%%%%%%%%%%%%%%%%%%%%%%%%%%%%%%%%%%%%%%%%%%%%%%%%%
%%%%%%%%%%%%%%%%%%%%%%%%%%%%%%%%%%%%%%%%%%%%%%%%%%%%%%%%%%%%%%%%%%%%
%%%%%%%%%%%%%%%%%%%%%%%%%%%%%%%%%%%%%%%%%%%%%%%%%%%%%%%%%%%%%%%%%%%%
	In this paper we will prove the existence of unique, essentially smooth distributional solutions 
to the quasi-linear elliptic partial differential problem of divergence form
\begin{alignat}{1}
\nabla \cdot \frac{ \nabla\,u (s)} {\sqrt{1 - | \nabla u(s) |^2 }}
+
   4 \pi \sum_{n=1}^N a_n \delta_{s_n}(s) 
& =0
\qquad \mbox{for}\qquad\; s\in\Rset^3,
 \label{eq:uPDE}\\
u(s)
&\to 0
\qquad \mbox{as}\qquad |s|\to\infty;
\label{eq:uASYMPnull}
\end{alignat}
here, $\delta_{s_n}$ is the unit Dirac measure supported at $s_n\in\Rset^3$, and
the $a_n\in \Rset\backslash\{0\}$ are amplitudes.
	More precisely, we will prove the following theorem:
\begin{Theo}\label{Thm:MIKI}
	For any finite sets $\{s_n\}_{n=1}^N\subset\Rset^3$ and 
$\{a_n\}_{n=1}^N\subset \Rset\backslash\{0\}$ there exists a unique real function
$u\in C^{0,1}(\Rset^3)$ which solves \Ref{eq:uPDE}, \Ref{eq:uASYMPnull} in the sense of
distributions.
	Furthermore, $|\nabla{u}(s)|<1$ for $s\in\Rset^3\backslash\{s_n\}_{n=1}^N$, 
and $\lim_{s\to s_n}|\nabla{u}(s)| = 1$ for each $s_n$.
	Thus, $u\in C^\omega(\Rset^3\backslash\{s_n\}_{n=1}^N)$.
\end{Theo}
\begin{Rema}
	Evidently our theorem allows that some of the support points for the Dirac measures coincide;
however, any such situation is identical to a reformulation of the problem with fewer but distinct points,
with a re-assignment of amplitude values.
	Thus, without loss of generality we will henceforth assume that all the $s_n$ are distinct.
	The amplitudes may or may not be distinct, though.
\end{Rema}
	Our result has applications in physics and geometry.
	It governs objects as diverse as, on the
one hand, the classical electrostatic fields of the Maxwell-Born-Infeld field theory 
\cite{BornInfeldBb, PryceB, GibbonsA, KieMBIinJSPa}, and maximal spacelike hypersurfaces with lightcone 
defects in the Minkowski spacetime \cite{BartnikSimon, Ecker, KlyMik, KlyA, KlyB} on the other.
	Applications are discussed in section 4.

	Curiously enough, given the attention that these areas of research have received in the literature, 
the existence of solutions to \Ref{eq:uPDE}, \Ref{eq:uASYMPnull} as ascertained in Theorem \ref{Thm:MIKI} has 
been an unsettled problem.
	Of course, there is the explicit solution of \Ref{eq:uPDE}, \Ref{eq:uASYMPnull} for $N=1$
found by Born \cite{BornA} and elaborated on further in 
\cite{BornB, BornInfeldBb, BartnikSimon, Ecker, GibbonsA}; 
it is well-defined for any value of its amplitude $a$.
	There is also a semi-explicit solution of \Ref{eq:uPDE} (which violates \Ref{eq:uASYMPnull}, though) for $N=\infty$ 
found by Hoppe \cite{HoppeA} and further elaborated on in \cite{HoppeB, GibbonsA};
it has positive and negative amplitude Dirac sources of magnitude $|a|$ arranged in a cubic lattice and
exists for arbitrary $|a|$.
	However, to the best of our knowledge, generic existence theorems for solutions to \Ref{eq:uPDE} have so far 
been established only: (a) in \cite{KlyMik, KlyA} under smallness conditions\footnote{We note that
		the review of Klyachin's paper \cite{KlyA} in Mathematical Reviews incorrectly claims
		that existence were proved for arbitrary amplitudes.}
for the $a_n$ when \Ref{eq:uPDE} is restricted to (bounded or unbounded) domains with boundary with Dirichlet data 
replacing \Ref{eq:uASYMPnull}; (b) in \cite{KlyB} for arbitrary $a_n$ but with \Ref{eq:uASYMPnull}
replaced by prescribing $u(s_n)=u_n$, restricted by the bounds $|u_n-u_m| < |s_n-s_m|$ for 
$1\leq n< m\leq N$ --- in this case  \Ref{eq:uASYMPnull} is generically violated, and it doesn't follow from the proof
in \cite{KlyB} whether  \Ref{eq:uASYMPnull} can hold for some particular choices of 
$\{u_n\}_{n=1}^N\subset \Rset$ and $\{a_n\}_{n=1}^N\subset \Rset\backslash\{0\}$, given 
$\{s_n\}_{n=1}^N\subset\Rset^3$.
	Our Theorem \ref{Thm:MIKI} does not follow from adapting the proofs in \cite{KlyMik}, \cite{KlyA}, or 
\cite{KlyB}.
	In fact, our arguments also extend to the Dirichlet problem in domains with boundary,
as will become clear from our proof.

	As do their proofs of their theorems in \cite{KlyMik}, \cite{KlyA}, and \cite{KlyB},
our proof of Theorem \ref{Thm:MIKI} makes convenient use of the results by Bartnik and Simon \cite{BartnikSimon}.
	Explicitly, in  \cite{BartnikSimon}  Bartnik and Simon prove a number of results for the Dirichlet problem
of \Ref{eq:uPDE} in bounded domains (with almost arbitrarily irregular boundary!), and 
they also outline how to pass to unbounded domains using barrier functions as in \cite{Treibergs}.
	{From} their results one can extract the following theorem:
\begin{Theo}\label{Thm:BS} (Essentially Bartnik--Simon.)
	For any finite set $\{s_n\}_{n=1}^N\subset\Rset^3$ of distinct points and any finite set
$\{u_n\}_{n=1}^N\subset \Rset$, restricted by the bounds 
\beq
\quad\ |u_n-u_m| < |s_n-s_m|\qquad \mbox{for}\qquad 1\leq n< m\leq N,
\eeq
there exists a unique real function $u\in C^{0,1}(\Rset^3)$ which weakly solves 
\begin{alignat}{3}
	\nabla \cdot \frac{ \nabla\,u (s)} {\sqrt{1 - | \nabla u(s) |^2 }}
& =0 \qquad 
&\mbox{for}\qquad\; s\;
&\in\,\Rset^3\backslash\{s_n\}_{n=1}^N,
 \label{eq:uPDEmax}\\
	u(s)
&\to u_n\qquad 
&\mbox{as}\;\qquad  s\;
&\to s_n,
 \label{eq:uASYMPinner}\\
	u(s)
&\to 0\qquad 
&\mbox{as}\qquad  |s|
&\to\infty.
 \label{eq:uASYMPouter}
\end{alignat}
	Furthermore, $|\nabla{u}(s)|<1$ for $s\in\Rset^3\backslash\{s_n\}_{n=1}^N$, and
$u\in C^\omega(\Rset^3\backslash\{s_n\}_{n=1}^N)$.
\end{Theo}
	Theorem \ref{Thm:BS} basically reduces the proof of our 
Theorem \ref{Thm:MIKI} to variational arguments which show that for each set of amplitudes 
$\{a_n\}_{n=1}^N\subset \Rset\backslash\{0\}$ associated with the points $\{s_n\}_{n=1}^N\subset\Rset^3$ 
there exists a unique distributional $C^{0,1}(\Rset^3)$ solution $u(s)$ of \Ref{eq:uPDE}, \Ref{eq:uASYMPnull} 
for which  $|u(s_n)-u(s_m)| < |s_n-s_m|$ for $1\leq n< m\leq N$.
	The claim that $\lim_{|s-s_n|\to 0}|\nabla{u}(s)| = 1$ then follows from:
\begin{Theo}\label{Thm:Ecker} (Rephrasing of Theorem 1.5 in \cite{Ecker}). Let $u(s)$ be as in Theorem \ref{Thm:BS}.
	Then either $u(s)$ can be analytically continued into $s_n$ or 
$$\lim_{|s-s_n|\to 0}|\nabla{u}(s)| = 1.$$ 
	Thus, genuine singularities of $u(s)$ are lightcone singularities. 
\end{Theo}
	In section 2 we formulate our variational approach to \Ref{eq:uPDE}, \Ref{eq:uASYMPnull} and prove existence 
of a unique optimizer $u\in C^{0,1}_0(\Rset^3)$ with $|\nabla u|\leq 1$. 
	In section 3 we show with a dual variational argument that $|u(s_n)-u(s_m)| < |s_n-s_m|$ for $1\leq n< m\leq N$.
	Afterwards, in section 4, we discuss applications to physics and geometry.
	In section 5 we list a few straightforward extensions of our main theorem, only indicating their proofs.
	In section 6 we conclude with a list of desirable extensions.

\newpage
%%%%%%%%%%%%%%%%%%%%%%%%%%%%%%%%%%%%%%%%%%%%%%%%%%%%%%%%%%%%%%%%%%%%
%%%%%%%%%%%%%%%%%%%%%%%%%%%%%%%%%%%%%%%%%%%%%%%%%%%%%%%%%%%%%%%%%%%%
%%%%%%%%%%%%%%%%%%%%%%%%%%%%%%%%%%%%%%%%%%%%%%%%%%%%%%%%%%%%%%%%%%%%
%%%%%%%%%%%%%%%%%%%%%%%%%%%%%%%%%%%%%%%%%%%%%%%%%%%%%%%%%%%%%%%%%%%%
  \section{The variational approach}
%%%%%%%%%%%%%%%%%%%%%%%%%%%%%%%%%%%%%%%%%%%%%%%%%%%%%%%%%%%%%%%%%%%%
%%%%%%%%%%%%%%%%%%%%%%%%%%%%%%%%%%%%%%%%%%%%%%%%%%%%%%%%%%%%%%%%%%%%
%%%%%%%%%%%%%%%%%%%%%%%%%%%%%%%%%%%%%%%%%%%%%%%%%%%%%%%%%%%%%%%%%%%%
%%%%%%%%%%%%%%%%%%%%%%%%%%%%%%%%%%%%%%%%%%%%%%%%%%%%%%%%%%%%%%%%%%%%
	In this section we prove:
\begin{Prop}\label{Prp:uVAR} There exists a unique
$u\in C^{0,1}_0(\Rset^3)\cap\{v:|\nabla{v}|\leq 1\}$ for which 
\beq
0
=
\int_{\Rset^3}
\Bigl(\nabla \psi(s)\cdot \frac{\nabla{u}(s)}{\sqrt{1 - \abs{\nabla u(s)}{}^2}}
	-  4 \pi \psi(s)\sum_{1\leq n\leq N} a_n \delta_{s_n}(s)\Bigr)\dd^3s
\label{eq:uPDEweak}
\eeq
holds for all $\psi\in C^\infty_0(\Rset^3)$, where $\nabla{u}$ denotes weak derivative and
where $\dd^3s$ is three-dimensional Lebesgue measure.
\end{Prop}
	Thus, $u\in C^{0,1}_0(\Rset^3)\cap\{v:|\nabla{v}|\leq 1\}$ is a distributional solution of
\Ref{eq:uPDE}, \Ref{eq:uASYMPnull}.
%%%%%%%%%%%%%%%%%%%%%%%%%%%%%%%%%%%%%%%%%%%%%%%%%%%%%%%%%%%%%%%%%%%%
%%%%%%%%%%%%%%%%%%%%%%%%%%%%%%%%%%%%%%%%%%%%%%%%%%%%%%%%%%%%%%%%%%%%
%%%%%%%%%%%%%%%%%%%%%%%%%%%%%%%%%%%%%%%%%%%%%%%%%%%%%%%%%%%%%%%%%%%%
	\subsection{Preliminary considerations}
%%%%%%%%%%%%%%%%%%%%%%%%%%%%%%%%%%%%%%%%%%%%%%%%%%%%%%%%%%%%%%%%%%%%
%%%%%%%%%%%%%%%%%%%%%%%%%%%%%%%%%%%%%%%%%%%%%%%%%%%%%%%%%%%%%%%%%%%%
%%%%%%%%%%%%%%%%%%%%%%%%%%%%%%%%%%%%%%%%%%%%%%%%%%%%%%%%%%%%%%%%%%%%
	Equation \Ref{eq:uPDE} is the formal Euler-Lagrange equation for the variational principle
\beq
	\cF(v) = \int_{\Rset^3} \Bigl(1- \sqrt{1 - \abs{\nabla v(s)}{}^2}
		-  4 \pi v(s)\sum_{1\leq n\leq N} a_n \delta_{s_n}(s)\Bigr)\dd^3s\; \to\;\mbox{min}\qquad
\label{eq:VPv}
\eeq
over a suitable set of functions $v$, and \Ref{eq:uPDEweak} is its weak version.
	In particular, if $C^0_b(\Rset^3)\cap C^1_b(\Rset^3\backslash\{s_n\}_{n=1}^N)$ denotes the Banach 
space of bounded continuous real functions on $\Rset^3$ which have a bounded continuous
derivative on the indicated punctured domain, equipped with their usual norm, then $\cF$ is 
well-defined for those 
$v\in C^0_b(\Rset^3)\cap C^1_b(\Rset^3\backslash\{s_n\}_{n=1}^N)$ which satisfy $|\nabla{v}|\leq 1$ on 
$\Rset^3\backslash\{s_n\}_{n=1}^N$ and which vanish sufficiently fast as $|s|\to\infty$; in particular, 
$|\nabla{v}(s)| = O(|s|^{-2})$ is fast enough.
	Eventually, in section 3, we will show that $\cF$ does take its minimizer on this set of functions.
	However, since the indicated Banach spaces are not convenient spaces to work with, 
here we shall characterize \Ref{eq:VPv} as upper limit of a sequence of 
variational functionals which are defined on larger, more convenient spaces of functions. 
	The minimizer of $\cF$ will be obtained as the limit of a family of minimizers of
these approximating variational principles.
	In particular, we show that the minimizer solves \Ref{eq:uPDEweak}.
%%%%%%%%%%%%%%%%%%%%%%%%%%%%%%%%%%%%%%%%%%%%%%%%%%%%%%%%%%%%%%%%%%%%
%%%%%%%%%%%%%%%%%%%%%%%%%%%%%%%%%%%%%%%%%%%%%%%%%%%%%%%%%%%%%%%%%%%%
%%%%%%%%%%%%%%%%%%%%%%%%%%%%%%%%%%%%%%%%%%%%%%%%%%%%%%%%%%%%%%%%%%%%
	\subsection{A monotone family of variational principles}
%%%%%%%%%%%%%%%%%%%%%%%%%%%%%%%%%%%%%%%%%%%%%%%%%%%%%%%%%%%%%%%%%%%%
%%%%%%%%%%%%%%%%%%%%%%%%%%%%%%%%%%%%%%%%%%%%%%%%%%%%%%%%%%%%%%%%%%%%
%%%%%%%%%%%%%%%%%%%%%%%%%%%%%%%%%%%%%%%%%%%%%%%%%%%%%%%%%%%%%%%%%%%%
	For $x\geq 0$ we define the extended real-valued function 
\beq
f(x) = \left\{ {1- \sqrt{1 - x}\qquad {\mathrm{for}} \qquad x\in[0,1]
                                  \atop\, 
                  \infty\qquad\qquad\ {\mathrm{for}} \qquad x >1}
       \right.
\eeq
	The $K$-th Taylor polynomial of $f$ about $x=0$, given by
\beq
\Tay_{K}[f](x|0)  = \sum_{k=0}^K f^{(k)}(0) x^{k},
\label{eq:TaylorPolK}
\eeq
has Taylor coefficients
\beq
f^{(k)}(0) = \left\{\begin{array}{lrl}
\quad{0}                &{\mathrm{for}} &k = 0\\
\quad\frac{1}{2}        &{\mathrm{for}} &k = 1\\
\frac{(2k-3)!!}{(2k)!!} &{\mathrm{for}} &k > 1
                     \end{array}
              \right.,
\label{eq:TaylorcK}
\eeq
so that $\Tay[f](x|0)\equiv \big\{\Tay_{K}[f](x|0)\big\}_{K=1}^\infty$, the Maclaurin series of $f(x)$, 
viz.
\beq
\Tay[f](x|0) =  \frac{1}{2}x + \sum_{k=2}^\infty \frac{(2k-3)!!}{(2k)!!}x^{k},
\label{eq:McL}
\eeq
is a pointwise strictly increasing sequence of strictly convex, strictly increasing functions of $x > 0$
which vanish at $x=0$.
	The series \Ref{eq:McL} converges absolutely to $f(x)$ for $x\in[0,1]$  but diverges for $x>1$;
since all coefficients are positive, $\Tay_{K}[f](x|0) \nearrow\infty$ for $x>1$, so we are
entitled to say that $\Tay[f](x|0)$ actually converges to the extended real-valued function $f(x)$ 
for all $x\geq 0$. 
	In the following, for brevity we shall simply write $f_{K}(x)$ for $\Tay_{K}[f](x|0)$.

	With the help of the Taylor polynomials we now define the family of functionals
\beq
	\cF_K(v) = \int_{\Rset^3} \Bigl(f_{K}\left(|\nabla{v}|^2\right)
		-  4 \pi v(s)\sum_{1\leq n\leq N} a_n \delta_{s_n}(s)\Bigr)\dd^3s,
\label{eq:VPvK}
\eeq
which  for $K\geq 2$ are well-defined\footnote{The functional $\cF_1$ is not well-defined on the 
	``canonical'' domain of the Dirichlet integral, which is $\dot{W}^{1,2}_0(\Rset^3)$, 
	for which reason we don't have any use for $\cF_K$ when $K=1$.} 
on $\bigcap_{1\leq k\leq K}\dot{W}_0^{1,2k}(\Rset^3)$.
	For the $|\nabla{v}|$ term this is seen right away from the definition of $\dot{W}_0^{1,2k}(\Rset^3)$ 
as the closure of the compactly supported $C^\infty$ functions on $\Rset^3$ w.r.t. 
$\norm{v}_{\dot{W}_0^{1,2k}(\Rset^3)}^{2k}\equiv \int_{\Rset^3}|\nabla{v}(s)|^{2k}\dd^3s$, 
so that for $v\in \bigcap_{1\leq k\leq K}\dot{W}_0^{1,2k}(\Rset^3)$ we have 
$f_{K}(|\nabla{v}|^2)\in L^1(\Rset^3)$. 
	To see that also the source term in \Ref{eq:VPvK} is well-defined we note that
$\dot{W}_0^{1,2k}(\Rset^3)\subset {W}_{\mathrm{loc}}^{1,2k}(\Rset^3)$ so that we can apply
Sobolev's original embedding theorem, according to which for any ball $B\subset \Rset^3$ we 
have\footnote{By the Sobolev-Morrey embedding theorem we even have
		${W}^{1,2k}(B)\hookrightarrow C^{0,1-3/2k}(B)$ for $k\geq 2$ and any
		$B\subset \Rset^3$.}
${W}^{1,2k}(B)\hookrightarrow C^{0}_b(B)$ whenever $k\geq 2$, and conclude that
$\delta_{s_n}\in \dot{W}_0^{-1,(2k)^\pr}(\Rset^3)$ for all $k\geq 2$.
	This establishes that $\cF_K$ is well-defined on $\bigcap_{1\leq k\leq K}\dot{W}_0^{1,2k}(\Rset^3)$
for $K\geq 2$, and any finite $N$.
	Incidentally, since elements of $\dot{W}_0^{1,2}(\Rset^3)$ tend to zero
at spatial infinity\footnote{This is not true for all elements of $\dot{W}_0^{1,2k}(\Rset^3)$ when $k\geq 2$}
a.e., we see that the $v\in \bigcap_{1\leq k\leq K}\dot{W}_0^{1,2k}(\Rset^3)$ with  $K\geq 2$ are in fact
equivalent to a subset of the bounded continuous functions on $\Rset^3$ which vanish at spatial infinity.

	Let ${\admA}\equiv\bigcap_{1\leq k\leq\infty}\dot{W}_0^{1,2k}(\Rset^3)\cap\{v:|\nabla{v}|\leq 1\}$ 
be the set of admissible (equivalence classes of) functions for $\cF$.
	Since ${\admA}\subset\bigcap_{1\leq k\leq K}\dot{W}_0^{1,2k}(\Rset^3)$ for all $K\in \Nset$, 
and since $f_K(x)\nearrow f(x)$ for all $x\geq 0$, monotone convergence now yields that 
$\cF(v)=\lim_{K\to\infty}\cF_K(v)$ for all $v\in\admA$.
%%%%%%%%%%%%%%%%%%%%%%%%%%%%%%%%%%%%%%%%%%%%%%%%%%%%%%%%%%%%%%%%%%%%
%%%%%%%%%%%%%%%%%%%%%%%%%%%%%%%%%%%%%%%%%%%%%%%%%%%%%%%%%%%%%%%%%%%%
%%%%%%%%%%%%%%%%%%%%%%%%%%%%%%%%%%%%%%%%%%%%%%%%%%%%%%%%%%%%%%%%%%%%
	\subsection{Existence of a family of unique minimizers}
%%%%%%%%%%%%%%%%%%%%%%%%%%%%%%%%%%%%%%%%%%%%%%%%%%%%%%%%%%%%%%%%%%%%
%%%%%%%%%%%%%%%%%%%%%%%%%%%%%%%%%%%%%%%%%%%%%%%%%%%%%%%%%%%%%%%%%%%%
%%%%%%%%%%%%%%%%%%%%%%%%%%%%%%%%%%%%%%%%%%%%%%%%%%%%%%%%%%%%%%%%%%%%
	For each $K\geq 2$, the functional $\cF_K$ is clearly convex over 
$\bigcap_{1\leq k\leq K}\dot{W}_0^{1,2k}(\Rset^3)$.
	Moreover, $\cF_K$ is bounded below and 
coercive w.r.t. the topology of $\bigcap_{1\leq k\leq K}\dot{W}_0^{1,2k}(\Rset^3)$ 
whenever $K\geq 2$.
	To see this we have to estimate the source term in $\cF_K$. 

	We rewrite $\cF_K$ as
\beq
	\cF_K(v) = 
\sum_{k=1}^K c_k\norm{v}_{\dot{W}^{1,2k}_0(\Rset^3)}^{2k}	-  4 \pi\sum_{1\leq n\leq N} a_n v(s_n),
\label{eq:VPvKnorm}
\eeq
where $c_k = f^{(k)}(0)>0$.
	Now $\{s_n\}_{n=1}^N$ is given, so there exists an open  ball $B\subset \Rset^3$ such that 
$\{s_n\}_{n=1}^N\subset B$. 
	Since the restriction of any $v\in \dot{W}_0^{1,4}(\Rset^3)$ to $B$ is in $\dot{W}^{1,4}(B)$, 
and since $\dot{W}^{1,4}(B)\subset {W}^{1,4}(B)$ (though no embedding, clearly), we can apply the Sobolev 
embedding theorem in the form ${W}^{1,4}(B)\hookrightarrow C^0_b(B)$,
and using $|a_n|\leq \max_{1\leq n\leq N}|a_n|<\infty$, for all $K\geq 2$ we obtain the estimate
\beq
    \cF_K(v) \geq \sum_{k=1}^K c_k\norm{v}_{\dot{W}^{1,2k}_0(\Rset^3)}^{2k} -  4 \pi N A\norm{v}_{W^{1,4}(B)}
\label{eq:VPvKprecoercivity}
\eeq
where $A$ is a positive constant. 
	Now $\norm{v}_{W^{1,4}(B)}^4 = \norm{v}_{L^4(B)}^4 + \norm{\nabla{v}}_{L^4(B)}^4$,  so that
$\norm{v}_{W^{1,4}(B)} \leq \norm{v}_{L^4(B)} + \norm{\nabla{v}}_{L^4(B)}$.
	Next, since 
$v\in \dot{W}_0^{1,4}(\Rset^3)$ for $K\geq 2$, we have the nontrivial estimate
$\norm{\nabla{v}}_{L^4(B)}\leq \norm{v}_{\dot{W}^{1,4}_0(\Rset^3)}<\infty$.
	Furthermore, by H\"older's inequality, $\norm{v}_{L^4(B)}\leq |B|^{1/12}\norm{v}_{L^6(B)}$,
and the special case $\dot{W}_0^{1,2}(\Rset^3)\hookrightarrow L^6(\Rset^3)$
of Sobolev's embedding theorem then yields the estimate
$\norm{v}_{L^4(B)}\leq |B|^{1/12}S \norm{v}_{\dot{W}^{1,2}_0(\Rset^3)}<\infty$, 
where $S>0$ is the sharp Sobolev constant.
	And so, for $K\geq 2$ we find 
\beq
	\cF_K(v) \geq
\sum_{k=1}^K c_k\norm{v}_{\dot{W}^{1,2k}_0(\Rset^3)}^{2k}
	-  4 \pi N \left(A'\norm{v}_{\dot{W}^{1,2}_0(\Rset^3)} + A\norm{v}_{\dot{W}^{1,4}_0(\Rset^3)}\right),
\label{eq:VPvKcoercivity}
\eeq
where $A'$ is another positive constant. 
	This lower estimate for $\cF_K$ is manifestly bounded below and coercive on
$\bigcap_{1\leq k\leq K}\dot{W}_0^{1,2k}(\Rset^3)$ whenever $K\geq 2$.
	Therefore, for each $K\geq2$ the functional $\cF_K$ takes on a unique minimum for some 
$v_K\in \bigcap_{1\leq k\leq K} \dot{W}_0^{1,2k}(\Rset^3)$.
	We set $\cF_K(v_K) \equiv F_K$. 
%%%%%%%%%%%%%%%%%%%%%%%%%%%%%%%%%%%%%%%%%%%%%%%%%%%%%%%%%%%%%%%%%%%%
%%%%%%%%%%%%%%%%%%%%%%%%%%%%%%%%%%%%%%%%%%%%%%%%%%%%%%%%%%%%%%%%%%%%
%%%%%%%%%%%%%%%%%%%%%%%%%%%%%%%%%%%%%%%%%%%%%%%%%%%%%%%%%%%%%%%%%%%%
	\subsection{Weak convergence of the family of minimizers}
%%%%%%%%%%%%%%%%%%%%%%%%%%%%%%%%%%%%%%%%%%%%%%%%%%%%%%%%%%%%%%%%%%%%
%%%%%%%%%%%%%%%%%%%%%%%%%%%%%%%%%%%%%%%%%%%%%%%%%%%%%%%%%%%%%%%%%%%%
%%%%%%%%%%%%%%%%%%%%%%%%%%%%%%%%%%%%%%%%%%%%%%%%%%%%%%%%%%%%%%%%%%%%
	Since $F_K=\cF_K(v_K) > \cF_{K^\pr}(v_K)\geq\cF_{K^\pr}(v_{K^\pr}) = F_{K^\pr}$ whenever 
$K>K^\pr$, the minimum values $F_K$ of the family of variational functionals form a strictly 
monotonic increasing sequence. 
	And since $f_K(x) < f(x)$ for all $K$ when $x>0$, this sequence $\{F_K\}_{K=2}^\infty$ is 
bounded above by $\cF(\hat{v})$, where $\hat{v}\in C^{0,1}(\Rset^3)$ is the following convenient 
trial function: 
let $2r := \min\{|s_k-s_l|\}_{1\leq k<l\leq N}$, then $\hat{v}$ is defined by
\beq
\hat{v}(s) = 
\left\{
\begin{array}{rll} 
\sign(a_n)\ (r- |s_n-s|) &{\mathrm{for}} &s\in B_r(s_n)\\
                       0 &{\mathrm{for}} &s\in\Rset^3\backslash\bigcup_{1\leq n\leq N} B_r(s_n)
\end{array}
\right.
\eeq
	One readily calculates that 
\beq
\cF(\hat{v})= N\Bigl(|B_r| - 4\pi r\, \overline{|a|}\Bigr),
\eeq
where 
\beq
\overline{|a|} \equiv \frac{1}{N} \sum_{1\leq n\leq N}|a_n|.
\eeq
	Thus, $\lim_{K\to\infty}F_K =:F \leq \inf_v\cF(v) \leq \cF(\hat{v})$ exists, 
and $F_K <F$ for all $K\geq 2$. 

	As a corollary, since $\cF_{K^\pr}(v_K) < \cF_K(v_K)$ when $K>K^\pr$, we have that
$\cF_{K^\pr}(v_K)<F$ whenever $K>K^\pr$.
	By coercivity, for any fixed $K^\pr \geq 2$ there now exists a positive constant $C$ 
such that $\norm{v_K}_{\dot{W}_0^{1,2K^\pr}(\Rset^3)}< CF$ for all $K>K^\pr$.
	Now, since $\dot{W}_0^{1,2K^\pr}(\Rset^3)$ is a separable, reflexive Banach space for all 
$1\leq K^\pr<\infty$, the closed ball 
$\bigl\{v: \norm{v}_{\dot{W}_0^{1,2K^\pr}(\Rset^3)}< CF\bigr\}$ is weakly compact. 
	Therefore the sequence $\{v_K\}_{K=2}^\infty$ contains a weakly convergent subsequence in 
$(\dot{W}_0^{1,2}\cap \dot{W}_0^{1,2K^\pr})(\Rset^3)$ for each $K^\pr \geq 1$.
	By a diagonal argument we can pick the subsequence so that its weak limit in each 
$(\dot{W}_0^{1,2}\cap \dot{W}_0^{1,2K^\pr})(\Rset^3)$ is one and the same 
$v_\infty\in \bigcap_{1\leq k\leq K^\pr} \dot{W}_0^{1,2k}(\Rset^3)$ for all $1\leq K^\pr< \infty$, 
hence $v_\infty\in \bigcap_{1\leq k<\infty} \dot{W}_0^{1,2k}(\Rset^3)$.

	Now, functions in $\bigcap_{1\leq k < \infty} \dot{W}_0^{1,2k}(\Rset^3)$ are not 
necessarily in $\dot{W}_0^{1,\infty}$, but the uniform (in $K$) upper bound $F$ on the 
$\cF_{K}(v_K)$ guarantees that the weak limit $v_\infty$ of the $v_K$ is actually in 
$\dot{W}_0^{1,\infty}$; indeed, we even have $|\nabla{v}_\infty|\leq 1$ a.e.
	For assume to the contrary that $|\nabla{v}_\infty|\not\leq 1$ a.e. 
	Then there exists an $\Omega\subset\Rset^3$ with $|\Omega|>0$ such that 
$|\nabla{v}_\infty|\geq 1+2\eps$ a.e. in $\Omega$. 
	But then there exists a $\widetilde{K}$ such that $|\nabla{v}_K|\geq 1+\eps$ a.e. in 
$\Omega$ whenever $K> \widetilde{K}$. 
	And then we have 
$\cF_K(v_K) 
\geq |\Omega| \sum_{k=\widetilde{K}+1}^K \frac{(2k-3)!!}{(2k)!!}(1+\eps)^{2k}\nearrow \infty$ 
as $K\to\infty$, which is a contradiction to $\cF_K(v_K)< F$ for all $K$. 
	Thus, not only is $v_\infty\in \dot{W}_0^{1,\infty}$, also $|\nabla{v}_\infty|\leq 1$ 
a.e. in $\Rset^3$, as claimed.
 So $v_\infty\in\bigcap_{1\leq k\leq\infty}\dot{W}_0^{1,2k}(\Rset^3)\cap\{v:|\nabla{v}|\leq 1\}=\admA$. 
%%%%%%%%%%%%%%%%%%%%%%%%%%%%%%%%%%%%%%%%%%%%%%%%%%%%%%%%%%%%%%%%%%%%
%%%%%%%%%%%%%%%%%%%%%%%%%%%%%%%%%%%%%%%%%%%%%%%%%%%%%%%%%%%%%%%%%%%%
%%%%%%%%%%%%%%%%%%%%%%%%%%%%%%%%%%%%%%%%%%%%%%%%%%%%%%%%%%%%%%%%%%%%
	\subsection{The limit of the minimizers of the $\cF_K$ minimizes $\cF$}
%%%%%%%%%%%%%%%%%%%%%%%%%%%%%%%%%%%%%%%%%%%%%%%%%%%%%%%%%%%%%%%%%%%%
%%%%%%%%%%%%%%%%%%%%%%%%%%%%%%%%%%%%%%%%%%%%%%%%%%%%%%%%%%%%%%%%%%%%
%%%%%%%%%%%%%%%%%%%%%%%%%%%%%%%%%%%%%%%%%%%%%%%%%%%%%%%%%%%%%%%%%%%%
	We have just proved that $v_\infty\in{\admA}$.  
	We now show that $\cF(v_\infty) = \min_{v\in\admA}\cF(v)$.

	Since ${\admA}\subset ({W}^{1,2}\cap{W}^{1,2K})(\Rset^3)$ for each $K\geq 1$, 
$\cF_K(v)$ is well-defined for each $v\in{\admA}$ and each $K\in\Nset$; in particular, 
$\cF_K(v_\infty)$ is well-defined for each $K\in\Nset$.
	Moreover, $\lim_{K\to\infty} \cF_K(v)=\cF(v)$ for each $v\in{\admA}$, by 
monotone convergence.

	Now, by the monotonicity of the sequence of Taylor polynomials
$\big\{f_{K}(|\nabla{v}|)\big\}_{K=1}^\infty$, we have 
$\cF(v_\infty) > \cF_K(v_\infty)\geq \cF_K(v_K) =F_K$ for all $K>1$.
	By taking the limit as $K\to\infty$, 
we obtain $\cF(v_\infty) \geq \lim_{K\to\infty}\cF_K(v_K) =F$.

	On the other hand, recalling \Ref{eq:VPvKnorm},
we see that  each $\cF_K$ is obviously weakly lower semi-continuous, so
we have $\cF_{K^\pr}(v_\infty)\leq \lim_{K\to\infty}\cF_{K^\pr}(v_K)$. 
	Recalling now that $\cF_{K^\pr}(v_K)<F$ whenever $K>K^\pr$, 
we obtain $\cF_{K^\pr}(v_\infty)\leq F$ for all $K^\pr>1$.
	Taking the limit $K^\pr\to\infty$ and recalling that
$\lim_{K^\pr\to\infty} \cF_{K^\pr}(v)=\cF(v)$ for each $v\in{\admA}$, we obtain that
$\cF(v_\infty) \leq F$.
	In total, we have shown that $\cF(v_\infty) = F$. 

	It remains to show that there is no $\tilde{v}\in {\admA}$ for which $\cF(\tilde{v})<F$. 
	But this is really easy. 
	For assume to the contrary that there were such a 
$\tilde{v}$ with $\cF(\tilde{v}) = F-\eps$.
	Then $\cF_K(\tilde{v}) < F-\eps$ for all $K>1$, which contradicts the fact that for
any $\eps$ we can find a $\widetilde{K}(\eps)$ such that  
$\min_{v\in({W}^{1,2}\cap{W}^{1,2K})(\Rset^3)}\cF_K(v) = \cF_K(v_K) > F-\eps$ whenever 
$K>\widetilde{K}(\eps)$.

   This proves that $\cF$ takes its  minimum at $v_\infty\in\admA$, and the minimum equals $F$.
   Moreover, by convexity, the minimizer is unique.
%%%%%%%%%%%%%%%%%%%%%%%%%%%%%%%%%%%%%%%%%%%%%%%%%%%%%%%%%%%%%%%%%%%%
%%%%%%%%%%%%%%%%%%%%%%%%%%%%%%%%%%%%%%%%%%%%%%%%%%%%%%%%%%%%%%%%%%%%
%%%%%%%%%%%%%%%%%%%%%%%%%%%%%%%%%%%%%%%%%%%%%%%%%%%%%%%%%%%%%%%%%%%%
	\subsection{The minimizer of $\cF$ weakly satisfies the Euler-Lagrange equation}
%%%%%%%%%%%%%%%%%%%%%%%%%%%%%%%%%%%%%%%%%%%%%%%%%%%%%%%%%%%%%%%%%%%%
%%%%%%%%%%%%%%%%%%%%%%%%%%%%%%%%%%%%%%%%%%%%%%%%%%%%%%%%%%%%%%%%%%%%
%%%%%%%%%%%%%%%%%%%%%%%%%%%%%%%%%%%%%%%%%%%%%%%%%%%%%%%%%%%%%%%%%%%%
	We cannot yet conclude that the minimizer $v_\infty$ of $\cF(v)$ weakly satisfies the 
formal Euler-Lagrange equation \Ref{eq:uPDE} because for this conclusion we need to know that 
$|\nabla{v}_\infty|<1$ a.e. and so far we only know that $|\nabla{v_\infty}|\leq 1$ a.e.
	We now show that $|\nabla{v}_\infty|<1$ a.e., which implies that $v_\infty$ 
weakly satisfies \Ref{eq:uPDE}, i.e. \Ref{eq:uPDEweak}.

	Let $\Omc= \bigcap_{\eps>0}\overline{\{s:|\nabla{v_\infty}|>1-\eps\}}$. 
	Note that $\Omc$ contains all points $s_*$ at which $|\nabla{v_\infty}(s_*)|=1$ as well as 
all points $s_*$ for which ess-$\lim_{s\to s_*} |\nabla{v_\infty}(s)|=1$ without necessarily
having $\nabla{v_\infty}(s)$ itself defined at $s=s_*$.
	Clearly, $\Omc$ has finite Lebesgue measure, $|\Omc|<\infty$, for ${v_\infty}\in{\admA}$
implies that $|\nabla{v_\infty}(s)|\to 0$ as $|s|\to\infty$. 
	We now show that $|\Omc|=0$.

	For this purpose we assume to the contrary that $|\Omc|>0$. 
	Then the variation of $\cF(v)$ about $v_\infty$ gives to the leading order (i.e. power 1/2) in $\psi$,
\beq
	\cF^{(1/2)}[v_\infty](\psi) 
:= 
- \int_{\Omc} \sqrt{- 2\nabla{v_\infty(s)}\cdot\nabla\psi(s)}\, \dd^3s,
\label{eq:VPvpsiHALF}
\eeq
where $\psi\in C^\infty_0(\Rset^3)$ is any test function satisfying 
$\nabla{v_\infty(s)}\cdot\nabla\psi(s) \leq 0$ a.e. on $\Omc$.
	Note that $\cF^{(1/2)}[v_\infty](\psi)$ is homogeneous of fractional degree 1/2 in $\psi$; 
hence, this \emph{nonlinear} term --- if nonzero --- will in general dominate the usual linear 
terms in $\psi$, indeed.
	Moreover, whenever $\cF^{(1/2)}[v_\infty](\psi) \neq 0$, we manifestly have
\beq
-\int_{\Omc} \sqrt{- \nabla{v_\infty(s)}\cdot\nabla\psi(s)}\,\dd^3s < 0.
\label{eq:VPvpsiNEG}
\eeq
	But for $v_\infty$ to minimize $\cF$ over $\admA$ we must have 
$\cF^{(1/2)}[v_\infty](\psi) \geq 0$ for all $\psi\in C^\infty_0(\Rset^3)$ satisfying 
$\nabla{v_\infty(s)}\cdot\nabla\psi(s) \leq 0$ on $\Omc$ a.e. 
	This together with \Ref{eq:VPvpsiNEG} implies that 
$\cF^{(1/2)}[v_\infty](\psi) \equiv 0$ for all 
$\psi\in C^\infty_0(\Rset^3)$ satisfying $\nabla{v_\infty(s)}\cdot\nabla\psi(s) \leq 0$ 
a.e. on $\Omc$.
	But this is only possible if $|\Omc|=0$, as claimed.

\begin{Rema}
	Our argument above does not show that $\Omc = \{s_n\}_{n=1}^N$.
\end{Rema}

	The result $|\Omc|=0$ means that $|\nabla v_\infty| <1$ a.e., and
this already implies that the variation of $\cF(v)$ about $v_\infty$ to leading order (i.e. power 1) in 
$\psi$ now reads
\beq
	\cF^{(1)}[v_\infty](\psi) 
= 
 \int_{\Rset^3}
 \Bigl(\nabla \psi(s)\cdot \frac{\nabla{v_\infty}(s)}{\sqrt{1 - \abs{\nabla v_\infty(s)}{}^2}}
	-  4 \pi \psi(s)\sum_{1\leq n\leq N} a_n \delta_{s_n}(s)\Bigr)\dd^3s.
\label{eq:VPvpsiONE}
\eeq
	Since $\cF^{(1)}[v_\infty](\psi)$ is linear in $\psi$, $v_\infty$ can minimize $\cF$ over 
$\admA$ only if $\cF^{(1)}[v_\infty](\psi) =0$ for all $\psi$, which is precisely \Ref{eq:uPDEweak}.
	Thus the Euler-Lagrange equation \Ref{eq:uPDE} is satisfied by $v_\infty$ in the weak 
sense, as claimed.
	The proof of Proposition \ref{Prp:uVAR} is complete.
\QED

\begin{Rema}
	We close this section with the remark that alternate, nonvariational routes to Proposition
\ref{Prp:uVAR} are conceivable.
	In particular, the Dirac sources can be mollified with compactly supported $C^\infty$ functions,
and the asymptotic vanishing of $u(s)$ as $|s|\to\infty$ replaced by $0$-Dirichlet conditions on $\partial B_R$,
where $R$ is a large ball containing the supports of all mollifiers of the Dirac sources.
	For this situation the theorems in \cite{BartnikSimon} guarantee a classical solution to the so
mollified \Ref{eq:uPDE}.
	As pointed out by one of the referees, Lemma 2.1 in \cite{BartnikSimon} and elliptic regularity theory 
should now yield uniform Lipschitz bounds on the solution $u$ away from the eventual locations of the Dirac sources 
when the mollifiers are removed, and the proof of their Lemma 3.1 shows that the limit function solves 
\Ref{eq:uPDEweak} restricted to $B_R$.
	Subsequently one can let $R\to\infty$ by invoking Treiberg's barrier function arguments.
\end{Rema}
%%%%%%%%%%%%%%%%%%%%%%%%%%%%%%%%%%%%%%%%%%%%%%%%%%%%%%%%%%%%%%%%%%%%
%%%%%%%%%%%%%%%%%%%%%%%%%%%%%%%%%%%%%%%%%%%%%%%%%%%%%%%%%%%%%%%%%%%%
%%%%%%%%%%%%%%%%%%%%%%%%%%%%%%%%%%%%%%%%%%%%%%%%%%%%%%%%%%%%%%%%%%%%
%%%%%%%%%%%%%%%%%%%%%%%%%%%%%%%%%%%%%%%%%%%%%%%%%%%%%%%%%%%%%%%%%%%%
	\section{Bootstrapping regularity}
%%%%%%%%%%%%%%%%%%%%%%%%%%%%%%%%%%%%%%%%%%%%%%%%%%%%%%%%%%%%%%%%%%%%
%%%%%%%%%%%%%%%%%%%%%%%%%%%%%%%%%%%%%%%%%%%%%%%%%%%%%%%%%%%%%%%%%%%%
%%%%%%%%%%%%%%%%%%%%%%%%%%%%%%%%%%%%%%%%%%%%%%%%%%%%%%%%%%%%%%%%%%%%
%%%%%%%%%%%%%%%%%%%%%%%%%%%%%%%%%%%%%%%%%%%%%%%%%%%%%%%%%%%%%%%%%%%%
	In this section we bootstrap the regularity of the minimizer $v_\infty\equiv u$ of $\cF(v)$ 
to the level which guarantees satisfaction of Theorem \ref{Thm:MIKI}.
%%%%%%%%%%%%%%%%%%%%%%%%%%%%%%%%%%%%%%%%%%%%%%%%%%%%%%%%%%%%%%%%%%%%
%%%%%%%%%%%%%%%%%%%%%%%%%%%%%%%%%%%%%%%%%%%%%%%%%%%%%%%%%%%%%%%%%%%%
%%%%%%%%%%%%%%%%%%%%%%%%%%%%%%%%%%%%%%%%%%%%%%%%%%%%%%%%%%%%%%%%%%%%
	\subsection{Bootstrapping the regularity of $u$ away from $\Omc$}
%%%%%%%%%%%%%%%%%%%%%%%%%%%%%%%%%%%%%%%%%%%%%%%%%%%%%%%%%%%%%%%%%%%%
%%%%%%%%%%%%%%%%%%%%%%%%%%%%%%%%%%%%%%%%%%%%%%%%%%%%%%%%%%%%%%%%%%%%
%%%%%%%%%%%%%%%%%%%%%%%%%%%%%%%%%%%%%%%%%%%%%%%%%%%%%%%%%%%%%%%%%%%%
	By our Proposition \ref{Prp:uVAR}, the unique distributional solution to \Ref{eq:uPDE}, \Ref{eq:uASYMPnull} 
obtained by minimizing $\cF$ in $\admA$ takes values $u_n=u(s_n)$ at the $s_n$ which satisfy the inequalities 
$|u_n-u_m| \leq |s_n-s_m|$ for all $1\leq n< m\leq N$.
	Hence we can invoke Corollary 4.2 to Theorem 4.1 of \cite{BartnikSimon} to extract the following 
proposition for our setting.
\begin{Prop}\label{Prp:BSb}
	Let $u(s)\in C^{0,1}(\Rset^3)$ be the unique distributional solution to \Ref{eq:uPDE}, \Ref{eq:uASYMPnull} 
which minimizes $\cF$ in $\admA$. 
	Then $u\in C^\omega(\Rset^3\backslash \Omc)$.
	Moreover, $\Omc$ (the singular set $K$ for $u$ in \cite{BartnikSimon}) is a subgraph of 
$\cK_N\equiv \cK\big(\{s_n\}_{n=1}^N \big)\subset \Rset^3$, the 
complete graph whose vertices are the set $\{s_n\}_{n=1}^N$. 
	Furthermore, let $E_{n,m}\subset \cK_N$ denote the edge of $\cK_N$ with endpoints $s_n$ and $s_m$.
	Then $E_{n,m}\subset \Omc$  if and only if $|u_n-u_m| = |s_n-s_m|$, and in that case we have
$u(ts_n+(1-t)s_m) = t u_n + (1-t)u_m$ for $t\in[0,1]$.
\end{Prop}
%%%%%%%%%%%%%%%%%%%%%%%%%%%%%%%%%%%%%%%%%%%%%%%%%%%%%%%%%%%%%%%%%%%%
%%%%%%%%%%%%%%%%%%%%%%%%%%%%%%%%%%%%%%%%%%%%%%%%%%%%%%%%%%%%%%%%%%%%
%%%%%%%%%%%%%%%%%%%%%%%%%%%%%%%%%%%%%%%%%%%%%%%%%%%%%%%%%%%%%%%%%%%%
	\subsection{Proof that $\Omc = \{s_n\}_{n=1}^N$}
%%%%%%%%%%%%%%%%%%%%%%%%%%%%%%%%%%%%%%%%%%%%%%%%%%%%%%%%%%%%%%%%%%%%
%%%%%%%%%%%%%%%%%%%%%%%%%%%%%%%%%%%%%%%%%%%%%%%%%%%%%%%%%%%%%%%%%%%%
%%%%%%%%%%%%%%%%%%%%%%%%%%%%%%%%%%%%%%%%%%%%%%%%%%%%%%%%%%%%%%%%%%%%
	We recall that any distributional solution $\in W^{1,2}$ of \Ref{eq:uPDE} satisfies the 
weak maximum principle, Theorem 8.1 in \cite{GilbargTrudinger}.
	Therefore $v_\infty(s)\equiv u(s)$ has a local maximum at $s_n$ whenever $a_n>0$ and a local 
minimum when $a_n<0$, and no extremum in $\Rset^3\backslash\{s_n\}_{n=1}^N$.
	This together with Proposition \ref{Prp:BSb} right away gives us:
\begin{Coro}\label{Cor:PPMMedgesAREout}
	Let $a_na_m >0$. Then $E_{nm}\backslash\{s_n,s_m\}\not\subset \Omc$.
\end{Coro}
	Thus, the only potentially critical edges $E_{n,m}$ are those whose end points $s_n$ and $s_m$ sport 
amplitudes $a_n$ and $a_m$ of different sign.
	To show that also those edges, save their endpoints, are not critical requires a different argument.
	We shall invoke a convex duality argument which rules out all the edges, save their endpoints, from
the critical set.
\begin{Prop}\label{Prp:PMedgesAREout}
	For all $1\leq n<m\leq N$ we have that $E_{nm}\backslash\{s_n,s_m\}\not\subset \Omc$.
\end{Prop}
\textit{Proof:}
	Since $C^\infty_0(\Rset^3)$ is dense in $C^{0,1}_0(\Rset^3)$, we can substitute $v_\infty=u$ for $\psi$ in
\Ref{eq:uPDEweak} and, for the solution $u$ of \Ref{eq:uPDEweak}, obtain the identity
\beq
0
=
\int_{\Rset^3}
\Bigl(\frac{|\nabla{u}(s)|^2}{\sqrt{1 - \abs{\nabla u(s)}{}^2}}
	-  4 \pi u(s)\sum_{1\leq n\leq N} a_n \delta_{s_n}(s)\Bigr)\dd^3s.
\label{eq:uPDEweakSELF}
\eeq
	A simple rewriting of \Ref{eq:uPDEweakSELF} yields 
\beq
\cF(u)
=
\int_{\Rset^3}
\Bigl(1 - \frac{1}{\sqrt{1 - \abs{\nabla u(s)}{}^2}}\Bigr)\dd^3s.
\label{eq:Fidentity}
\eeq
	Defining
\beq
U(s)
=
\frac{-\nabla{u}(s)}{\sqrt{1 - \abs{\nabla u(s)}{}^2}},
\label{eq:uV}
\eeq
where $u$ is still the solution of \Ref{eq:uPDEweak}, another elementary rewriting of \Ref{eq:Fidentity}
yields that $\cF(u) = - \cG (U)$, where
\begin{equation}
\hskip-5pt
  \cG (V)
= \label{eq:FIELDenergyMBI}
\int_{\mathbb R^3}\!
\Bigl[\!{\sqrt{1 + |{V}|^2 }} -1\Bigr] \dd^3s
\end{equation}
is well-defined for any vector field $V$ for which $|V|\in \Lloc^1(\Rset^3)\cap L^2(\Rset^3\backslash B_R)$,
where $B_R$ is some ball of large radius $R$.
	Note that $\cG(V)$ is related to $\cF(v)$ by a Legendre--Fenchel transform, viz.
\begin{equation} 
\hskip-5pt
\cG (V)  
= \label{eq:LEGENDREtransf}
\max_{v\in C^{0,1}_0} \int_{\mathbb R^3}\!\left(
	\Bigl[\!{\sqrt{1 - |{\nabla v}|^2 }}-1\Bigr] - V \cdot\nabla v\right)\dd^3s;
\end{equation}
the dual variables of the transformation are $\nabla v\leftrightarrow V$.
	Thus, $\cG (V)$ is strictly convex in $V$.
	But we have seen that also $\cF(v)$ is strictly convex for $v\in \admA$, so $\cF(v)$ --- or rather the 
source-free part of $\cF(v)$ --- is given as a Legendre--Fenchel transform of $\cG(V)$.
	As a result, we can also obtain the minimum of $\cF(v)$ and its minimizer $v_\infty =u$ in 
terms of a constrained minimum principle for $\cG(U)$.
	Explicitly, 
\begin{equation} 
\hskip-5pt
\cG (U)  
= \label{eq:dualVAR}
\min \Big\{\cG(V)\Big| \nabla\cdot V 
= 4\pi\textstyle\sum\limits_{n=1}^N a_n \delta_{s_n}; \; |V|\in \Lloc^1(\Rset^3)\cap L^2(\Rset^3\backslash B_R)\Big\};
\end{equation}
in \Ref{eq:dualVAR} it is understood that $\nabla\cdot V$ is well-defined in the sense of distributions and that
$R$ is big enough so that $\{s_n\}_{n=1}^N\subset B_R$.
\newpage

	Next, we define the almost everywhere harmonic field
\begin{equation} 
\hskip-5pt
V_h(s)  
= \label{eq:fundHARMfield}
 - {\textstyle\sum\limits_{n=1}^N} a_n \nabla \frac{1}{|s-s_n|}.
\end{equation}
	Note that $V_h(s) \in \Lloc^1(\Rset^3)\cap L^2(\Rset^3\backslash B_R)$ whenever 
$\{s_n\}_{n=1}^N\subset B_R$.
	Furthermore,
\begin{equation} 
\nabla\cdot V_h
= \label{eq:fundHARMlaw}
4\pi\textstyle\sum\limits_{n=1}^N a_n \delta_{s_n}.
\end{equation}
	So $V_h$ is in the set of admissible vector fields for our variational principle \Ref{eq:dualVAR}.

	We are now ready for our main argument.
	Namely, suppose that for some $n,m$ the edge $E_{n,m}\subset\Omc$.
	Without loss of generality we may assume that no other $s_k$ lies on $E_{n,m}$.
	Then $\lim_{s\to E_{n,m}} |\nabla u(s)|=1$, and so $\lim_{s\to E_{n,m}} |U(s)|=\infty$.
	But since $u$ is analytic away from $\Omc$, so is $U$, hence we conclude that there is some tubular 
neighborhood of $E_{n,m}$ in which $|U(s)|> |V_h(s)|$. 
	Since $s_n\neq s_m$ we can intersect our tubular neighborhood of $E_{n,m}$ with two small balls centered
at $s_n$ and $s_m$, respectively, and delete the intersection domain from it. 
	Denote the resulting truncated tubular neighborhood by $E^\circ_{n,m}$; it is a bounded open set. 
	Mollify its boundary $\partial E^\circ_{n,m}$ a little bit to obtain an open corridor $C^\circ_{n,m}$;
it needs to have a finite distance from any $s_k$.
	Now let $V_*(s)$ be given by $U(s)$ for $s\not\in E^\circ_{n,m}\cup C^\circ_{n,m}$, and by 
$V_*(s) = V_h(s)$ for $s\in E^\circ_{n,m}\backslash C^\circ_{n,m}$.
	We need to connect these fields smoothly across $C^\circ_{n,m}$, but this is easy.
	Since away from $\{s_n\}_{n=1}^N$ the fields $U$ and $V_h$ are divergence-free, we can represent each
field as the curl of some vector field.
	We can choose a $C^\infty$ deformation of one such vector field into the other
across the transition region  $C^\circ_{n,m}$, and in $C^\circ_{n,m}$ we define $V_*$ to be the curl of this 
deformed field.
	Thus $V_*$ is in the set of admissible vector fields for our variational principle.
	But then we have $\cG(V_*) < \cG(U)$, which is a contradition to our variational principle \Ref{eq:dualVAR}.

	Thus $|U(s)|< \infty$ for $s\in\Rset\backslash\{s_n\}_{n=1}^N$, and therefore $\Omc =\{s_n\}_{n=1}^N$.
\QED

\begin{Rema}
	We close this section with the remark that our convex duality argument can also be adapted to show that
$|\nabla v_\infty(s)|<1$ away from $\{s_n\}_{n=1}^N$ without invoking Proposition \ref{Prp:BSb}. 
	But then a Nash-Moser argument has to be supplied to bootstrap the regularity of $v_\infty$ from Lipschitz 
continuity to real analyticity in $\Rset^3\backslash\{s_n\}_{n=1}^N$. 
\end{Rema}
\vskip-1truecm $\phantom{nix}$
%%%%%%%%%%%%%%%%%%%%%%%%%%%%%%%%%%%%%%%%%%%%%%%%%%%%%%%%%%%%%%%%%%%%
%%%%%%%%%%%%%%%%%%%%%%%%%%%%%%%%%%%%%%%%%%%%%%%%%%%%%%%%%%%%%%%%%%%%
%%%%%%%%%%%%%%%%%%%%%%%%%%%%%%%%%%%%%%%%%%%%%%%%%%%%%%%%%%%%%%%%%%%%
%%%%%%%%%%%%%%%%%%%%%%%%%%%%%%%%%%%%%%%%%%%%%%%%%%%%%%%%%%%%%%%%%%%%
	\section{Applications to Geometry and Physics}
%%%%%%%%%%%%%%%%%%%%%%%%%%%%%%%%%%%%%%%%%%%%%%%%%%%%%%%%%%%%%%%%%%%%
%%%%%%%%%%%%%%%%%%%%%%%%%%%%%%%%%%%%%%%%%%%%%%%%%%%%%%%%%%%%%%%%%%%%
%%%%%%%%%%%%%%%%%%%%%%%%%%%%%%%%%%%%%%%%%%%%%%%%%%%%%%%%%%%%%%%%%%%%
%%%%%%%%%%%%%%%%%%%%%%%%%%%%%%%%%%%%%%%%%%%%%%%%%%%%%%%%%%%%%%%%%%%%
%
%%%%%%%%%%%%%%%%%%%%%%%%%%%%%%%%%%%%%%%%%%%%%%%%%%%%%%%%%%%%%%%%%%%%
%%%%%%%%%%%%%%%%%%%%%%%%%%%%%%%%%%%%%%%%%%%%%%%%%%%%%%%%%%%%%%%%%%%%
%%%%%%%%%%%%%%%%%%%%%%%%%%%%%%%%%%%%%%%%%%%%%%%%%%%%%%%%%%%%%%%%%%%%
	\subsection{Spacetime interpretation  of Theorem \ref{Thm:MIKI}.}
%%%%%%%%%%%%%%%%%%%%%%%%%%%%%%%%%%%%%%%%%%%%%%%%%%%%%%%%%%%%%%%%%%%%
%%%%%%%%%%%%%%%%%%%%%%%%%%%%%%%%%%%%%%%%%%%%%%%%%%%%%%%%%%%%%%%%%%%%
%%%%%%%%%%%%%%%%%%%%%%%%%%%%%%%%%%%%%%%%%%%%%%%%%%%%%%%%%%%%%%%%%%%%
	A smooth space-like hypersurface $\Sigma$ in Minkowski spacetime 
$\Mset^4\cong\Rset\times\Rset^3$ is a three-dimensional simply connected subset 
of $\Mset^4$ with a time-like normal vector at every point in $\Sigma$.
	Thus $\Sigma =\{\varpi\in\Mset^4: \cT(\varpi) =0\}$ is the boundary of the zero
level set of a differentiable function $\cT:\Mset^4\to\Rset$ with ran$(\cT)=\Rset$ and
$\dQ \cT(\varpi)$ time-like, i.e. $\gQ^{-1}(\dQ \cT(\varpi),\dQ \cT(\varpi))<0$ 
for all $\varpi\in \Mset^4$; here $\dQ$ is E. Cartan's exterior derivative on ${\Mset}^4$
and $\gQ$ the Minkowski metric with signature $+2$, a 2-covariant tensor acting on 
$T(\Mset^4)\times T(\Mset^4)$, where $T(\Mset^4)$ is the tangent bundle of $\Mset^4$.
	Topologically, $\Sigma\sim\Rset^3$. 
\newpage

	Since any such hypersurface is a graph over $\Rset^3$, without loss of generality we 
may assume that the generating function $T$ is of the form $T(\varpi)=t - c^{-1}S(s)$, where 
$\varpi\cong(ct,s)$ defines a Lorentz frame, where $t$ is time and $s$ is a vector in Euclidean space $\Rset^3$. 
	Then $\Sigma = \{(ct,s): t = c^{-1}S(s)\}$. 
	Since $\gQ^{-1}(\dQ T(\varpi),\dQ T(\varpi)) = -1 + \abs{\nabla S}^2$, 
and since $\Sigma$ is space-like, we need to have $1 - \abs{\nabla S}^2>0$ everywhere.

	For those $\Sigma$ which are asymptotically flat, more precisely if 
$\Sigma\asymp\Sigma_0$ with $\Sigma_0 \cong \Rset^3$, the volume difference 
$\triangle \mathrm{vol}(\Sigma|\Sigma_0)$ of $\Sigma$ versus $\Sigma_0$ is well-defined.
	After at most a Lorentz transformation we can choose $\Sigma_0 = \{(ct,s): t = 0\}\cong \Rset^3$, 
in which case

\vskip-10pt $\phantom{nix}$
\beq
	\triangle  \mathrm{vol}(\Sigma|\Sigma_0) = \int_{\Rset^3} \left(\sqrt{1 - \abs{\nabla S}^2} - 1\right)\dd^3s.
\label{eq:DIFFvol}
\eeq

	Note that $\triangle  \mathrm{vol}(\Sigma|\Sigma_0)\leq 0$.
	A hypersurface $\Sigma$ is called \emph{maximal} if any compact variation leads to a decrease of volume.
	In particular, $\Sigma_0\cong\Rset^3$ is a maximal entire spacelike hypersurface in $\Mset^4$.
	By a Bernstein theorem of Cheng and Yau	\cite{ChengYau}, \emph{any} entire space-like maximal 
hypersurface in $\Mset^4$ is flat; see also \cite{Yisong}.
	Thus, to be interesting a maximal hypersurface cannot be entirely space-like but
at best only space-like almost everywhere.
	If in particular $\Sigma$ has isolated defects then by Ecker's theorem these are lightcone singularities, 
i.e. isolated points in $\Sigma$ where the normal vector touches the lightcone.
	Any such almost-everywhere space-like maximal hypersurface with point defects in $\Mset^4$ is the graph
$\Sigma = \{(ct,s): t = c^{-1}S(s)\}$ of a function $S(s)$ satisfying $1 - \abs{\nabla S}^2>0$ away from the
defects, such that $1 - \abs{\nabla S}^2$  extends continuously into the defects, where it vanishes.
	
	Prescribing the locations $s_k\in \Rset^3$ of the lightcone singularities does not uniquely 
determine an asymptotically flat maximal hypersurface with defects.
	In addition, the particular asymptotically linear behavior of $S(s)$, and also 
the integral mean curvatures $\mu_k\in\Rset\backslash\{0\}$ which are associated with each lighcone singularity 
of the hypersurface have to be prescribed.
	Maximizing $\triangle \mathrm{vol}(\Sigma|\Sigma_0)$ for such a hypersurface with lightcone singularities
of prescribed integral mean curvatures is a variational problem with constraints.
	The Euler-Lagrange equation for this problem reads\footnote{We follow the convention 
		of \cite{GilbargTrudinger} which differs from that in \cite{BartnikSimon} by the factor $3$.} 
\beq
-\nabla \cdot
\frac{ \nabla\,S}{\sqrt{1 - |\nabla S|^2 }}
= 
   3\sum_{k=1}^N \mu_k \delta_{{s_k}}.
\label{eq:PDEforS}
\eeq 
	Identifying $S =u$ and $\mu_k= (4\pi/3) a_k$ yields \Ref{eq:uPDE}.
	
	In this spacetime interpretation our Theorem \ref{Thm:MIKI} becomes:
\begin{Coro}
	For any set $\{s_k\}_{k=1}^N\subset\Rset^3$ of distinct points and any set
of integral mean curvatures $\{\mu_k\}_{k=1}^N\subset\Rset\backslash\{0\}$ assigned to these points, 
there exists a unique asymptotically flat hypersurface $\Sigma = \{(ct,s): t = c^{-1}S(s)\}$ with
$S\in C^{0,1}_0(\Rset^3,\Rset)\cap C^\omega(\Rset^3\backslash\{{s_k}\}_{k=1}^N,\Rset)$ solving \Ref{eq:PDEforS};
moreover, $|\nabla S(s)|\to 1$ as $s\to s_k$.
	Thus $\Sigma$ is an almost everywhere space-like maximal hypersurface, having
$N$ lightcone singularities with prescribed integral mean curvatures $\mu_k$ located at the $s_k$.
\end{Coro}
%%%%%%%%%%%%%%%%%%%%%%%%%%%%%%%%%%%%%%%%%%%%%%%%%%%%%%%%%%%%%%%%%%%%
%%%%%%%%%%%%%%%%%%%%%%%%%%%%%%%%%%%%%%%%%%%%%%%%%%%%%%%%%%%%%%%%%%%%
%%%%%%%%%%%%%%%%%%%%%%%%%%%%%%%%%%%%%%%%%%%%%%%%%%%%%%%%%%%%%%%%%%%%
	\subsection{Electrostatic interpretation of Theorem \ref{Thm:MIKI}.}
%%%%%%%%%%%%%%%%%%%%%%%%%%%%%%%%%%%%%%%%%%%%%%%%%%%%%%%%%%%%%%%%%%%%
%%%%%%%%%%%%%%%%%%%%%%%%%%%%%%%%%%%%%%%%%%%%%%%%%%%%%%%%%%%%%%%%%%%%
%%%%%%%%%%%%%%%%%%%%%%%%%%%%%%%%%%%%%%%%%%%%%%%%%%%%%%%%%%%%%%%%%%%%
	In classical electromagnetic field theory, the Coulomb law states that an electric (point-)charge 
``density'' in $\Rset^3$ is the source of the electric displacement field $\SPvec{D}$,
\beq
        {\nabla}\cdot {\SPvec{D}}
=\label{eq:COULOMBlaw}
         4 \pi {\sum_{k=1}^N} \pmk\delta_{s_{k}}
	\phantom{blablabla}\mathrm{Coulomb's\ law}\,,
\eeq
with\footnote{Empirically, all nuclear and electron charges are integer multiples of the
		elementary charge, which is unity in our units.}
$\pmk\in \Zset\backslash\{0\}$, while Faraday's law says that an electrostatic field ${\SPvec{E}}$ is curl-free,
\beq
	\nabla\times\SPvec{E} 
=\label{eq:FARADAYlawSTATIC}
	 \SPvec{0} 
	\phantom{blablablablablabla} \mathrm{Faraday's\ law\ (stationary)} \,.
\eeq
	These two laws need to be complemented by a law relate ${\SPvec{E}}$ and ${\SPvec{D}}$, for which 
Max Born \cite{BornA} proposed
\beq
	{\SPvec{D}} 
= \label{eq:BORNlaw}
	\frac{{\SPvec{E}}}{ \sqrt{ 1 - \beta^4 |{\SPvec{E}}|^2  }} 
	\phantom{blablabla}\mathrm{Born's\ law} \phantom{blablablablab}
\eeq
with $\beta\in (0,\infty)$ (we use the dimensionless notation of \cite{KieMBIinJSPa}).
	In the limit $\beta\to 0$ Born's law \Ref{eq:BORNlaw} goes over into Maxwell's law of 
the ``pure aether'', ${\SPvec{D}}={\SPvec{E}}$.
	We remark that \Ref{eq:BORNlaw} is the electrostatic limit of both,
the electromagnetic law proposed by Born \cite{BornA, BornD} and the more elaborate 
law proposed by Born and Infeld \cite{BornInfeldBa, BornInfeldBb}. 
	The latter has received much attention in recent years, see the references in
\cite{GibbonsA, KieMBIinJSPa, KieMBIinJSPb, KieEMBIwDEFECTS}.

	Clearly, \Ref{eq:FARADAYlawSTATIC} implies that ${\SPvec{E}} = -\nabla A$ 
for some scalar potential field $A$.
	Inserting this representation for ${\SPvec{E}}$ into \Ref{eq:BORNlaw}, which in 
turn is inserted in  \Ref{eq:COULOMBlaw}, we obtain
\beq
    -    {\nabla}\cdot \frac{\nabla A}{ \sqrt{ 1 - \beta^4 |\nabla A|^2  }}
=
         4 \pi {\sum_{k=1}^N} \pmk\delta_{s_{k}}
\,.
\label{eq:PDEforA}
\eeq
	Multiplying \Ref{eq:PDEforA} by $\beta^2$ and identifying $\beta^2 A =u$ and 
$\beta^2\pmk=a_k$ we retrieve \Ref{eq:uPDE}.

	In this electrostatic interpretation our Theorem \ref{Thm:MIKI} yields:
\begin{Coro}\label{Cor:MBI}
	For any finite number $N$ of point charges $\{z_k\}_{k=1}^N\subset\Zset\backslash\{0\}$ located
at distinct points $\{s_k\}_{k=1}^N\subset\Rset^3$, there exists a unique electrostatic field ${\SPvec{E}}$ 
in $\Rset^3\backslash\{{s_k}\}_{k=1}^N$ which solves \Ref{eq:COULOMBlaw}, \Ref{eq:FARADAYlawSTATIC}, \Ref{eq:BORNlaw}
and has finite field energy\footnote{Here, $\alpha$ is Sommerfeld's fine structure constant, 
			inherited from the units in \cite{KieMBIinJSPa}.}
\beq
\cE_{\mathrm{field}}
\left({\SPvec{D}}\right) 
= 
\frac{1}{4\pi}
\frac{\alpha}{\beta^4}
\int_{\Rset^3}\left( \sqrt{ 1 + \beta^4|{\SPvec{D}}|^2 } - 1\right) \dd^3s.
\label{eq:HfuncFIELDS}
\eeq
	The solution ${\SPvec{E}} \in C^\omega (\Rset^3\backslash\{{s_k}\}_{k=1}^N,\Rset^3)$,
but it cannot be continuously extended into the ${s_k}$.
	It is bounded, with $\beta^{2}|\SPvec{E}(s)|\to 1$ for $s\to s_n$, and it vanishes for $|s|\to\infty$.
\end{Coro}
\newpage

\begin{Rema}
	Presumably inspired by Theorem 4.1 and Corollary 4.2 in \cite{BartnikSimon},
at the beginning of section 4 in \cite{GibbonsA}\footnote{In the same paragraph in
		\cite{GibbonsA} the results of Bartnik and Simon \cite{BartnikSimon} 
		(Gibbons' reference \cite{BartnikB}) for the weak solvability of the Dirichlet problem
		are somewhat misquoted: the necessary condition (in our notation) $|u_n-u_m|\leq |s_n-s_m|$ on the 
		Dirichlet data at the distinct points $s_n$ and $s_m$, which in Gibbons' notation would have to read 
		$|\Phi^a-\Phi^b|\leq |\xV_a-\xV_b|$, is missing.
		Whenever $|\Phi^a-\Phi^b| = |\xV_a-\xV_b|$, then a strut between $\xV_a$ and $\xV_b$ does occur.}
Gibbons contemplates the following: ``It is well known that one can construct explicit multi-black hole solutions 
held apart by struts, the struts being the sites of conical singularities representing
distributional stresses. One should be able to construct analogous multi-BIon solutions.'' 
	(What Gibbons calls ``multi-BIon'' solutions are but electrostatic solutions to the Maxwell--Born--Infeld 
equations with many point charge sources. 
	In particular, Born's solution, for Gibbons, is ``the BIon.'')
	Our Theorem \ref{Thm:MIKI} and its Corollary \ref{Cor:MBI} show that struts between the point charges
do not occur in the electrostatic solutions to the Maxwell--Born--Infeld field equations with point charge sources. 
\end{Rema}
%%%%%%%%%%%%%%%%%%%%%%%%%%%%%%%%%%%%%%%%%%%%%%%%%%%%%%%%%%%%%%%%%%%%
%%%%%%%%%%%%%%%%%%%%%%%%%%%%%%%%%%%%%%%%%%%%%%%%%%%%%%%%%%%%%%%%%%%%
%%%%%%%%%%%%%%%%%%%%%%%%%%%%%%%%%%%%%%%%%%%%%%%%%%%%%%%%%%%%%%%%%%%%
%%%%%%%%%%%%%%%%%%%%%%%%%%%%%%%%%%%%%%%%%%%%%%%%%%%%%%%%%%%%%%%%%%%%
	\section{Extensions of our results}
%%%%%%%%%%%%%%%%%%%%%%%%%%%%%%%%%%%%%%%%%%%%%%%%%%%%%%%%%%%%%%%%%%%%
%%%%%%%%%%%%%%%%%%%%%%%%%%%%%%%%%%%%%%%%%%%%%%%%%%%%%%%%%%%%%%%%%%%%
%%%%%%%%%%%%%%%%%%%%%%%%%%%%%%%%%%%%%%%%%%%%%%%%%%%%%%%%%%%%%%%%%%%%
%%%%%%%%%%%%%%%%%%%%%%%%%%%%%%%%%%%%%%%%%%%%%%%%%%%%%%%%%%%%%%%%%%%%
	The geometric interpretation of $u$ as time function of a maximal almost-everywhere space-like hypersurface
with lightcone singularities in Minkowski spacetime allows one to exploit the Poincar\'e group of $\Mset^4$
to generate solutions $u$ to \Ref{eq:uPDE} with different linear asymptotics at space-like infinity than 
\Ref{eq:uASYMPnull}.
	Since there is a unique Poincar\'e transformation for the transition from asymptotically vanishing to 
non-zero asymptotically linear conditions, we can therefore conclude:
\begin{Coro}\label{Cor:MBIcapacitor}
	For any finite sets $\{s_n\}_{n=1}^N\subset\Rset^3$ and $\{a_n\}_{n=1}^N\subset \Rset\backslash\{0\}$ 
and a vector $e\in\Rset^3$ of magnitude $|e|<1$ there exists a unique real function $u\in C^{0,1}(\Rset^3)$ 
satisfying
\begin{alignat}{1}
u(s) - e\cdot s \to 0
\qquad \mbox{as}\qquad |s|\to\infty,
\label{eq:uASYMPflat}
\end{alignat}
and solving \Ref{eq:uPDE} in the sense of distributions.
	Furthermore, $|\nabla{u}(s)|<1$ for $s\in\Rset^3\backslash\{s_n\}_{n=1}^N$, 
and $\lim_{s\to s_n}|\nabla{u}(s)| = 1$ for each $s_n$.
	Thus, $u\in C^\omega(\Rset^3\backslash\{s_n\}_{n=1}^N)$.
\end{Coro}
	The equivalence between the mathematical theories of maximal space-like 
hypersurfaces with point defects in Minkowski spacetime on the one hand and the electrostatic Maxwell--Born-Infeld 
potentials generated by point charge sources on the other allows us furthermore to re-interpret these asymptotically
nontrivially linear hypersurfaces as electrostatic solutions with asymptotically (at spacelike infinity) constant
electric fields.
	It is worth stressing that, in the notation of our previous subsection, one thus interprets
$(\beta^2 A, s)$, rather than the spacetime point $(ct,s)$, as Minkowski four-vector to generate new solutions by
Poincar\'e transformations.
	This ``hidden Poincar\'e symmetry'' seems to have been exploited first by Gibbons, 
see sections 3.3 and 3.7 of \cite{GibbonsA}; in particular, in section 3.7 Gibbons transforms Born's solution 
into an electrostatic solution with a single point charge and an asymptotically constant electric field whose 
magnitude is below Born's critical field strength.

	Lastly, as already announced in the introduction, there is an analogue of our Theorem \ref{Thm:MIKI} 
for the Dirichlet problem in bounded domains with nice boundary.
	This is not directly a corollary of our proof, yet its proof follows by a straightforward adaptation
of our proof to the Dirichlet problem.
	Thus we claim:
\begin{Theo}\label{Thm:DirichletMBI}
	Let $\Omega\subset\Rset^3$ be a bounded Lipschitz domain containing the finite point set
$\{s_n\}_{n=1}^N\subset\Rset^3$.
	Then for any $\phi:\partial\Omega\to\Rset$ satisfying $|\phi(s)-\phi(s')|< |s-s'|$ for $s\neq s'$,
and any set $\{a_n\}_{n=1}^N\subset \Rset\backslash\{0\}$, there exists a unique real $u\in C^{0,1}(\overline\Omega)$ 
solving
\begin{alignat}{1}
\nabla \cdot \frac{ \nabla\,u (s)} {\sqrt{1 - | \nabla u(s) |^2 }}
+
   4 \pi \sum_{n=1}^N a_n \delta_{s_n}(s) 
& =\; 0
\qquad\quad \mbox{for}\qquad\; s\in\Omega,
 \label{eq:uPDEagain}\\
u(s) 
&= \phi(s)
\qquad \mbox{for}\qquad s\in\partial\Omega
\label{eq:uDIRICHLETbc}
\end{alignat}
in the sense of distributions.
	Furthermore, $|\nabla{u}(s)|<1$ for $s\in\Omega\backslash\{s_n\}_{n=1}^N$, 
and $\lim_{s\to s_n}|\nabla{u}(s)| = 1$ for each $s_n$.
	Thus, $u\in C^\omega(\Omega\backslash\{s_n\}_{n=1}^N)$.
\end{Theo}
\begin{Rema}
	Bernd Kawohl kindly explained to me that for such a bounded domain the detour via the 
$\cF_K(v)$ should be unnecessary to minimize the convex functional $\cF(v)$ over the convex set 
$\{v\in W^{1,\infty}_0(\Omega):|\nabla v|\leq 1\, a.e.\ in\; \Omega\}$.
\end{Rema}
%%%%%%%%%%%%%%%%%%%%%%%%%%%%%%%%%%%%%%%%%%%%%%%%%%%%%%%%%%%%%%%%%%%%
%%%%%%%%%%%%%%%%%%%%%%%%%%%%%%%%%%%%%%%%%%%%%%%%%%%%%%%%%%%%%%%%%%%%
%%%%%%%%%%%%%%%%%%%%%%%%%%%%%%%%%%%%%%%%%%%%%%%%%%%%%%%%%%%%%%%%%%%%
%%%%%%%%%%%%%%%%%%%%%%%%%%%%%%%%%%%%%%%%%%%%%%%%%%%%%%%%%%%%%%%%%%%%
	\section{Desiderata}
%%%%%%%%%%%%%%%%%%%%%%%%%%%%%%%%%%%%%%%%%%%%%%%%%%%%%%%%%%%%%%%%%%%%
%%%%%%%%%%%%%%%%%%%%%%%%%%%%%%%%%%%%%%%%%%%%%%%%%%%%%%%%%%%%%%%%%%%%
%%%%%%%%%%%%%%%%%%%%%%%%%%%%%%%%%%%%%%%%%%%%%%%%%%%%%%%%%%%%%%%%%%%%
%%%%%%%%%%%%%%%%%%%%%%%%%%%%%%%%%%%%%%%%%%%%%%%%%%%%%%%%%%%%%%%%%%%%
	For matters of a quantitative nature it is important to have efficient algorithms to actually 
compute the solutions which in this paper we have proved do exist.
	For H\"older-continuous regularizations of the point charge sources such an algorithm has been developed
in \cite{CarKieSIAM, KieMBIinJMP}, but so far none seems available which would 
cover point charge and other singular sources in $\Rset^3$.
	The situation is better for the lower-dimensional problem in $\Rset^2$, see \cite{PryceA, Kobayashi, Ferraro},
and it is desirable also for the solutions in $\Rset^3$ to have explicit formulas in terms of, say, quadratures 
and such. 
	For the time being, the variational arguments allow one to work with numerical discretizations and 
to run minimization routines. 

	Another question is whether our Minkowski space results 
extend to certain curved Lorentz manifolds, in particular to asymptotically flat Lorentz manifolds
\cite{BartnikA}.
	If the Lorentz manifold is given (a so-called background spacetime), then 
the essence of the results of \cite{BartnikSimon} remains true under appropriate conditions, 
as shown by Bartnik in \cite{BartnikC} with quite different arguments than those in \cite{BartnikSimon}.
	Moreover, in \cite{BartnikD} Bartnik also extended Ecker's singularity theorem to certain Lorentz manifolds.
	For those  Lorentz manifolds for which the analogue of the flat spacetime theorems of 
Bartnik--Simon hold we expect that analogues of our theorems will hold as well. 
	We remark that Bartnik's theorems in \cite{BartnikC,BartnikD} do not require the manifold to be asymptotically 
flat.

	Another question, of prime importance as explained in \cite{KieEMBIwDEFECTS}, is whether the extension 
of our electrostatic results to a general-relativistic setting is possible in which an asymptotically flat 
Lorentz manifold is to be found by solving Einstein's field equations, with an electrostatic 
energy(-density)-momentum(-density)-stress tensor as curvature source for the metric, along with solving
the Maxwell--Born--Infeld equations for the electrostatic field in the curved spacetime.
	The problem with a single point charge source was treated already by Hoffmann \cite{HoffmannABC}
but only recently with complete rigor, by Tahvildar-Zadeh \cite{Shadi}; there the reader is also directed to the
large literature on the subject.
	The general sentiment, as expressed in the quote from Gibbons at the end of section 4, seems to
be that in the multi-point-charge problem struts will occur between the point charges; see also \cite{GilbertW}.
	We hope to offer a definitive answer in the foreseeable future.

\bigskip\noindent
\textbf{ACKNOWLEDGEMENT:} The work reported here has been funded by  NSF grant DMS-0807705.  
	NSF regulations require the disclaimer that: ``Any opinions, findings, and conclusions 
or recommendations expressed in this publication are those of the author and do not necessarily 
reflect the views of the National Science Foundation.'' 
	I gratefully acknowledge many insightful comments by my colleague Shadi Tahvildar-Zadeh.
	Many thanks go also to Bernd Kawohl and the two anonymous referees for their interesting
comments.
	Lastly I thank Penny Smith for many interesting discussions about her work \cite{Smith}
at an early stage of this project.

%%%%%%%%%%%%%%%%%%%%%%%%%%%%%%%%%%%%%%%%%%%%%%%%%%%%%%%%%%%%%%%%%%%%%
		%%%%%%%%   BIBLIOGRAPHY  %%%%%%%%%%%

\end{document}

\title{{\emph{Erratum}:}\\
ON THE QUASI-LINEAR ELLIPTIC PDE\\
      $-\nabla\cdot(\nabla{u}/\sqrt{1-|\nabla{u}|^2}) = 4\pi\sum_k a_k \delta_{s_k}$\\
	IN PHYSICS AND GEOMETRY\\
\small{Commun. Math. Phys. \textbf{314}:509--523 (2012).}}
\author{\sc{ Michael K.-H. Kiessling}\\
	Department of Mathematics\\
	Rutgers, The State University of New Jersey\\
	110 Frelinghuysen Rd., Piscataway, NJ 08854}
\date{Version of August 06, 2018; original Oct. 27, 2015}

\maketitle

\noindent
	Denis Bonheure kindly informed me that [1], section 2.6 (``\emph{The minimizer of $\cF$ weakly
satisfies the Euler--Lagrange equation}''), contains a logical gap on p.516~where~I~wrote:

	``The result $|\Omc|=0$ means that $|\nabla v_\infty| <1\ a.e.$, and
this already implies that the variation of $\cF(v)$ about $v_\infty$ to leading order (i.e. power 1) in 
$\psi$ now reads
\beq{\setcounter{equation}{22}}
	\cF^{(1)}[v_\infty](\psi) 
= 
 \int_{\Rset^3}
 \Bigl(\nabla \psi(s)\cdot \frac{\nabla{v_\infty}(s)}{\sqrt{1 - \abs{\nabla v_\infty(s)}{}^2}}
	-  4 \pi \psi(s){\textstyle\sum\limits_{\;1\leq n\leq N}} a_n \delta_{s_n}(s)\Bigr)\dd^3s.
\label{eq:VPvpsiONE}
\eeq
	Since $\cF^{(1)}[v_\infty](\psi)$ is linear in $\psi$, $v_\infty$ can minimize $\cF$ over 
$\admA$ only if $\cF^{(1)}[v_\infty](\psi) =0$ for all $\psi$, which is precisely (7). 
	Thus the Euler-Lagrange equation (1)  
is satisfied by $v_\infty$ in the weak sense, as claimed.''

      Bonheure's objection concerns the sentence: ``Since $\cF^{(1)}[v_\infty](\psi)$ is linear in $\psi$, ...'',
which alludes to the usual linearity-based argument (i.e., ``Suppose $\cF^{(1)}[v_\infty](\psi) \neq 0$ for some 
$\psi$; then either $\cF^{(1)}[v_\infty](\psi) < 0$ or $\cF^{(1)}[v_\infty](- \psi) < 0$; but this is impossible 
because $v_\infty$ is the minimizer of $\cF(v)$; hence, $\cF^{(1)}[v_\infty](\psi) = 0$.'').
      He notes that, although $|\Omc|=0$ (cf. the proof on p.515) implies that $\cF^{(1)}[v_\infty](\psi)$ 
is given by (22), 
and although (22) does act linearly on the space of compactly supported $C^\infty(\Rset^3)$ test functions $C^\infty_c(\Rset^3)$, 
only a nonlinear subset of these supplies admissible perturbations of $v_\infty$.
      More precisely, the restriction $v_\infty +\psi \in \admA$ (the admissible set of $v$) rules out 
test functions $\psi$ for which $\nabla{v_\infty(s)}\cdot\nabla\psi(s) \geq 0\ a.e.$ in some open $\epsilon$-ball 
$B_\epsilon$ satisfying
$\Omc\cap B_\epsilon\neq \emptyset$ while allowing those $\psi$
for which $\nabla{v_\infty(s)}\cdot\nabla\psi(s) < 0\ a.e.$ in such $B_\epsilon.$

\smallskip

\hrule
\smallskip
\noindent
\copyright{2017} {\small The author. Reproduction in its entirety is permitted for noncommercial purposes.}

\noindent
      Thus one cannot invoke the linearity of $\cF^{(1)}[v_\infty](\;\cdot\;):C^\infty_c(\Rset^3)\to\Rset$ 
to conclude that $\cF^{(1)}[v_\infty](\psi) = 0$ for all $\psi\in C^\infty_c(\Rset^3)$, which is equivalent to
the Euler--Lagrange equation (1) (see \eqref{eq:ELeqn} below) in weak form (see (7) in [1]).
   Of course, the nonlinear set of admissible $\psi$ contains the linear subset of 
$C^\infty_c(\Rset^3)$ test functions for which $\lim_{s\to\Omc} |\nabla\psi(s)| =0$ (incidentally, the 
larger linear subset of $C^\infty_c(\Rset)$ for which
$\lim_{s\to\Omc} \nabla\psi(s)\cdot\nabla{v_\infty}(s) =0$ contains inadmissible $\psi$, namely those for which
$\lim_{s\to\Omc} |\nabla\psi(s)| \neq 0$).
 For $\psi$ in this linear subset we do have $\cF^{(1)}[v_\infty](\psi) = 0$,
by the familiar argument; however, Bonheure points out,
knowing only that $|\Omc|=0$ does not allow us to conclude that the linear subset of test functions
satisfying $\lim_{s\to\Omc} |\nabla\psi(s)| =0$ is dense in $C^\infty_c(\Rset^3)$.

       In the following I respond to Bonheure's criticism (see also \cite{BoDAPoCPM}) by 
showing that $\cF^{(1)}[v_\infty](\psi) = 0$ not only for all $\psi\in C^\infty_c(\Rset^3)$
which satisfy $\lim_{s\to\Omc} |\nabla\psi(s)| = 0$, but indeed for all $\psi\in C^\infty_c(\Rset^3)$.
       I pick up on Remark 3.4, see p. 518 in [1]; thus we play a variation of the convex duality theme on p.517 of [1].

 We work with the almost everywhere harmonic field (eq.(29) in [1])
\begin{equation}\tag{E1}
\hskip-5pt
V_h(s)  
= \label{eq:fundHARMfield}
 - {\sum_{n=1}^N} a_n \nabla \frac{1}{|s-s_n|}.
\end{equation}
	We have (eq.(30) in [1])
\begin{equation}\tag{E2}
\nabla\cdot V_h 
=  \label{eq:COULOMBlaw}
4\pi\textstyle\sum\limits_{n=1}^N a_n \delta_{s_n}
\end{equation}
in the sense of distributions.
	Moreover, recall that for any
$w\in(\dot{W}_0^{1,p}(\Rset^3))^3$, $p\geq 1$,
\begin{equation}\tag{E3}
\nabla\cdot  \nabla\times w
=  \label{eq:divcurlISnull}
0
\end{equation}
weakly; note that $\nabla\times w$ is well defined on $\Rset^3$ except on a set $\Omw$ with Lebesgue measure zero.
 More generally, linearity implies that $\nabla\cdot  \nabla\times w =0$ 
for any $w=\sum_p w_p$ with weak curls 
$\nabla\times w_p \in ({L}^{p}(\Rset^3))^3$, where $\sum_p$ sums over a countable (sub-)set of $p\geq 1$.
 We will work with $\sum_p(\dot{W}_0^{\curl,p}(\Rset^3))^3$, $p\geq 1$, where
$(\dot{W}_0^{\curl,p}(\Rset^3))^3$ is defined as the closure of the set of divergence-free,
compactly supported $C^\infty$ vector fields $w$ with respect to the norm $\norm{\abs{\nabla\times w}}_{L^p(\Rset^3)}$.

          Abbreviating $V_h + \nabla\times w =:V$, an integration by parts now yields
\beq\nonumber
\begin{array}{ll}
	\cF(v_\infty)
 &= \displaystyle\int_{\Rset^3} \Bigl(1- \sqrt{1 - \abs{\nabla v_\infty(s)}{}^2}
		-  4 \pi v_\infty(s){\textstyle\sum\limits_{1\leq n\leq N}} a_n \delta_{s_n}(s)\Bigr)\dd^3s\\
{} &= \displaystyle\int_{\Rset^3} \Bigl(1- \sqrt{1 - \abs{\nabla v_\infty(s)}{}^2}
		+ V(s)\cdot  \nabla v_\infty(s)\Bigr)\dd^3s
\tag{E4}
\label{eq:cFminID}
\end{array}
\eeq  
for any such $w \in \sum_p(\dot{W}_0^{\curl,p}(\Rset^3))^3$, $p\geq 1$.
  Next observe that \emph{pointwise} 
\beq
\begin{array}{rl}\hspace{-10pt}
V(s)\cdot  \nabla v_\infty(s) -\sqrt{1- \abs{\nabla v_\infty(s)}{}^2}  \!\!&\geq 
\displaystyle\min_{E(s)\in B_1 a.e.}\!\!\Big\{\!\! -V(s)\cdot  E(s)-\sqrt{1 - \abs{E(s)}{}^2} \Big\}\!\! \tag{E5} \\
& 
= -\sqrt{1 + \abs{V(s)}{}^2_{}}, \quad s\in \Rset^3\setminus(\Omc\cup\Omw),\!\!\!
\end{array}
\eeq
where $B_1\subset \Rset^3$ is the open unit ball.
 The unique minimizer $E_V(s)$ is given by 
\beq\tag{E6}\label{eq:EsubVdef}
E_V(s) =  \frac{V(s)}{\sqrt{1 + \abs{V(s)}{}^2_{}}},
\eeq
defining a vector field on $\Rset^3\ a.e.$, satisfying $|E_V(s)|<1$, with $|E_V(s)|\to1$ when 
${s\to\{s_n\}_{n=1}^N}\subset\Omc$ and possibly when $s\to\Omw$.
 Inverting \eqref{eq:EsubVdef} yields
\beq\tag{E7}\label{eq:EsubVreverse}
V(s) = \frac{E_V(s)}{\sqrt{1 - \abs{E_V(s)}{}^2_{}}},
\eeq
and so, since  $\nabla\cdot V = 4\pi\textstyle\sum_{n=1}^N a_n \delta_{s_n}$ in the sense of distributions, we have
\begin{equation}\tag{E8}\label{eq:divVasperEsubV}
  \nabla\cdot  \frac{E_V(s)}{\sqrt{1 - \abs{E_V(s)}{}^2_{}}}
= 
4\pi\textstyle\sum\limits_{n=1}^N a_n \delta_{s_n}
\end{equation}
weakly, for \emph{any} $V=V_h+\nabla\times w$.
 This almost is the Euler--Lagrange equation we seek to obtain, yet not quite: at this point we don't know
whether $s\mapsto E_V(s)$ is a gradient field --- indeed, for most $w$ it's not! 
 Also, the minimization w.r.t. $E$ implies
\beq\tag{E9}
	\cF(v_\infty)
 \geq \displaystyle\int_{\Rset^3} \Bigl(1- \sqrt{1 + \abs{V(s)}{}^2}\Bigr)\dd^3s \equiv -\cG(V)
\label{eq:cFinfABOVEnegcG}
\eeq
for any $V=V_h+\nabla\times w$ with $w\in \sum_p(\dot{W}_0^{\curl,p}(\Rset^3))^3$, $p\geq 1$.
 So we need to show that there does exist a  
$U= V_h +\nabla\times w_*$ such that for $a.e.\ s\in\Rset^3$ we have $E_{U}(s) = -\nabla v_*(s)$ 
for some $v_*\in \dot{W}_0^{1,\infty}$, and with $\cG(U) = -\cF(v_\infty)$.
  The existence and uniqueness of the minimizer $v_\infty(s)$ of $\cF(v)$ (see [1])
then yields $v_*(s) = v_\infty(s)$. 
 Clearly, $U= V_h +\nabla\times w_*$ will minimize $\cG(V)$ amongst all $V=V_h+\nabla\times w$
with $w\in\sum_p(\dot{W}_0^{\curl,p}(\Rset^3))^3$, $p\geq 1$.   
 In fact, it suffices to minimize $\cG(V_h+\nabla\times w)$ for 
$w\in\sum_p(\dot{W}_0^{\curl,p}(\Rset^3))^3$, $p\in\{1,2\}$.
 We remark that $\sum_{p\in\{1,2\}}^{}(\dot{W}_0^{\curl,p}(\Rset^3))^3$ is a Banach space 
with respect to the norm
$\|w\|:=\inf\{\norm{\abs{\nabla\times w_1}}_{L^1(\Rset^3)} + \norm{\abs{\nabla\times w_2}}_{L^2(\Rset^3)}\}$,
where (given $w$) the infimum is over the set 
$\{w_1+w_2 =w\; |\; w_p\in (\dot{W}_0^{\curl,p}(\Rset^3))^3, p\in\{1,2\} \}$;
note that the splitting of $w$ into a sum of $w_1$ and $w_2$ is not unique (we will take advantage of this to prove Lemma 
\ref{lem:weakLtwoCMPCTness}, and Theorem \ref{theo:minEXISTS}, below).

 We now show that such a $w_*$ exists.

 In the special case $N=1$ it is easily seen that $w_*\equiv 0$ is the minimizer.
 Indeed, $V_h$ in this case is a spherically symmetric gradient field, and so is $E_{V_h}^{}$; 
thus, taking the Gateaux derivative $\frac{d}{dt}\cG(V_h+t\nabla\times w)$ at $t=0$ with compactly
supported $w\in C^\infty_c(\Rset^3)^3$ yields $\int\! E_{V_h}\cdot\nabla\times w\, d^3s$, and 
integration by parts now shows that this integral does vanish because $\nabla\times E_{V_h}=0$.
 Hence when $N=1$ then $V_h$ is a critical point of $\cG(V)$,
and the strict convexity of $\cG(V)$ w.r.t. $\nabla\times w$ now establishes its minimality.
 Of course, this just re-expresses the long-ago solved $\cF$ variational problem for $N=1$ in terms of the 
$\cG$ variational problem.
 Thus, in the following we assume $N>1$.
 
 When $N>1$ then $w\equiv 0$ is not a minimizer.
 For suppose $w\equiv 0$ were a minimizer, then the Gateaux derivative $\frac{d}{dt}\cG(V_h+t\nabla\times w)$ at $t=0$
would have to vanish, yet it is easily seen that it doesn't vanish for all $w$ 
because $\nabla\times E_{V_h}\not\equiv 0$ when $N>1$.
\newpage

 We next prove
\begin{Theo}\label{theo:minEXISTS}
$\exists! w_*\in (\dot{W}_0^{\curl,1}(\Rset^3))^3+(\dot{W}_0^{\curl,2}(\Rset^3))^3$ 
so that $U= V_h +\nabla\times w_*$ satisfies
\begin{equation} \tag{E10}
\hskip-5pt
\cG (U)  
= \label{eq:dualVP}
 \inf \Big\{\cG(V_h +\nabla\times w)\Big| \; w\in (\dot{W}_0^{\curl,1}(\Rset^3))^3+(\dot{W}_0^{\curl,2}(\Rset^3))^3\Big\}.
\end{equation}
\end{Theo}
\smallskip

\noindent
\textit{Proof of the Theorem}: 

 We begin by showing that $\cG(V)$ is well-defined on the stipulated set. 

\begin{Prop}\label{prop:wellDEFstrgCONT}
For $V=V_h+\nabla\times w$ with $w=w_1+w_2$ as stipulated, the functional $\cG(V)$ is well-defined and strongly continuous.
\end{Prop}

\noindent
\textit{Proof}: 

 Through telescoping
$\cG(V) = [\cG(V_h + \nabla\times w_1 + \nabla\times w_2) -\cG(V_h~+~\nabla~\times~w_1 ) ]
       + [\cG(V_h + \nabla\times w_1) -\cG(V_h)] +\cG(V_h)$, we right away
note that $\cG(V_h)$ is well-defined because 
$|V_h|\in \Lloc^1(\Rset^3)\cap L^2(\Rset^3\backslash B_R)$ for any open $B_R\supset \{s_n\}_{n=1}^N$, whereas
the two difference terms (the $[...]$ terms) are estimated as follows:
 We use the identity
\begin{equation}\tag{E11}\label{eq:cGvminuscGVh}
\cG(V_1 + V_2) -\cG(V_1)  
= \int_0^1 \int_{\Rset^3}
\frac{V_1(s) + \lambda V_2(s)}
{\sqrt{1+\abs{V_1(s) + \lambda V_2(s)}{}^2}}\cdot V_2(s)\dd^3s \dd\lambda,
\end{equation}
and note that $E_V=V\big/\sqrt{1+|V|{}^2}\in ((L^2\cap L^\infty)(\Rset^3))^3$, with $\norm{\abs{E_{V}}}_{L^\infty(\Rset^3)}=1$
and 
\begin{equation}\tag{E12}\label{eq:ABScGvminuscGVhBOUNDc}
\hspace{-10pt}
\norm{\abs{E_{V}}}_{L^2(\Rset^3)}
\leq 
|\Omega|^{1/2} + \norm{\abs{V_h}}_{L^2(\Rset^3\backslash \Omega)}+
\norm{\abs{\nabla\times w_1}}_{L^1(\Rset^3)}^{1/2} + \norm{\abs{\nabla\times w_2}}_{L^2(\Rset^3)},
\end{equation}
where $\Omega = B_R \cup \{\abs{\nabla\times w_1}> 1\}\cup \{\abs{\nabla\times w_2}> 1\}$
for some convenient $B_R\supset\{s_n\}_{n=1}^N$  (see appendix A); note that $\Omega$ is measurable but not
necessarily open.
 Thus, and setting $\norm{\;\cdot\;}_{L^p}^{}:= \norm{\;\cdot\;}_{L^p(\Rset^3)}^{}$ (below with $p=1$ or $2$),

\noindent
(i) let $V_1 = V_h$ and $V_2=\nabla\times w_1$, then H\"older's inequality applied to
\eqref{eq:cGvminuscGVh} yields
\begin{equation}\tag{E13}\label{eq:ABScGvminuscGVhBOUNDa}
\abs{\cG(V_h+\nabla\times w_1) - \cG(V_h)} \leq  \norm{\abs{\nabla\times w_1}}_{L^1};
\end{equation}

\noindent
(ii)  let $V_1 = V_h + \nabla\times w_1$ and $V_2=\nabla\times w_2$ and apply H\"older
to \eqref{eq:cGvminuscGVh} to get
\begin{equation}\tag{E14}\label{eq:ABScGvminuscGVhBOUNDb}
\abs{\cG(V_h+\nabla\times(w_1 + w_2)) - \cG(V_h+\nabla\times w_1)} 
\leq 
\sup_{0<\lambda<1} \norm{\abs{E_{V_\lambda}}}_{L^2} \norm{\abs{\nabla\times w_2}}_{L^2},
\end{equation}
where we have set $V_\lambda = V_h + \nabla\times w_1 +\lambda \nabla\times w_2$.
 The $L^2$ norm of $E_{V_\lambda}$ is estimated by \eqref{eq:ABScGvminuscGVhBOUNDc} with $w_2$ replaced by $\lambda w_2$, 
and the obvious estimate $\lambda< 1$.
 This establishes that  $\cG(V)$ is well-defined on the stipulated set. 

 This also proves that $\cG(V)$ is strongly continuous at $V_h$, for
$\big\|\big|\nabla\times w_1^{(n)}\big|\big\|_{L^1} \to 0$ and $\big\|\big|\nabla\times w_2^{(n)}\big|\big\|_{L^2} \to 0$
together imply $\cG(V_h +\nabla\times( w_1^{(n)} + w_2^{(n)})) \to \cG(V_h)$ as $n\to\infty$.
 The strong continuity of $\cG(V)$ at any $V= V_h +\nabla\times w$ with 
$w \in (\dot{W}_0^{\curl,1}(\Rset^3))^3+(\dot{W}_0^{\curl,2}(\Rset^3))^3$ follows essentially verbatim.
\hfill \QED
\newpage

 The strong continuity of $\cG(V)$ in concert with its strict convexity in $\nabla\times w$ implies:
\begin{Coro}\label{coro:wlsc}
The functional $\cG(V)$ is weakly lower semi-continuous.
\end{Coro}

\smallskip

 Since $\cG(V)$ is invariant under gauge transformations $w\to w + \nabla \gamma$ (because $\nabla\times w$, 
and thus $V$, are; also $E_V$ is gauge invariant), the strict convexity of $\cG(V)$ in $\nabla\times w$ does not
automatically translate into strict convexity of $\cG(V)$ in $w$.
 However, since we have stipulated $w$ to be divergence-free, viz. $\nabla\cdot w=0$, only 
gauge transformations $w\to w + \nabla \gamma$ with harmonic $\gamma$, i.e. with $\Delta\gamma =0$, remain; 
but the only allowed harmonic $\gamma$ are those which are constant at spatial $\infty$, which leaves the identity 
map as the only gauge transformation. 
 Thus $\cG(V)$ is strictly convex in $w$, and so we have 
\begin{Coro}\label{coro:uniqueness}
 Any minimizer $w_*\in (\dot{W}_0^{\curl,1}(\Rset^3))^3+(\dot{W}_0^{\curl,2}(\Rset^3))^3$ of $\cG(V_h+\curl w)$ 
is unique.
\end{Coro}

\smallskip

 We next show that
it suffices to consider $\cG(V_h+\nabla\times w)$ for $w\in (\dot{W}_0^{\curl,1}(\Rset^3))^3+(\dot{W}_0^{\curl,2}(\Rset^3))^3$.
 For this we recall the chiral Helmholtz theorem of \cite{Moses}:
\begin{Lemm}\label{lemm:Moses}
 Any vector field $\mathbf{F}(s)$ can be decomposed into $\mathbf{F}(s)=\mathbf{f}(s)+\mathbf{g}(s)$, where
$\mathbf{f}(s)$ is  divergence-free and $\mathbf{g}(s)$ is curl-free. 
 The divergence-free part $\mathbf{f}(s)$ has the chiral Fourier representation
\begin{equation}\tag{E15}\label{eq:MOSESrep}
\mathbf{f}(s) ={\textstyle\sum\limits_{\eta\in\{\pm1\}}}\int_{\Rset^3} e^{i2\pi k\cdot s}\mathbf{Q}_\eta(k) f_\eta(k) \dd^3k
\end{equation}
with $k=(k_1,k_2,k_3)^T$ and
$\mathbf{Q}_\eta(k)
=-\frac{\eta}{\sqrt{2}}\big(\frac{k_1(k_1+i\eta k_2)}{|k|(|k|+k_3)} -1, 
\frac{k_2(k_1+i\eta k_2)}{|k|(|k|+k_3)} -i\eta, \frac{k_1+i\eta k_2)}{|k|}\big)^T$.
 So $\mathbf{f}(s)$ is uniquely characterized by two scalar functions, $f_\eta(k)$, $\eta\in\{\pm1\}$, given by
\begin{equation}\tag{E16}\label{eq:MOSESrepINV}
f_\eta(k) = \int_{\Rset^3} e^{-i2\pi k\cdot s}\mathbf{Q}^*_\eta(k)\cdot \mathbf{F}(s) \dd^3s. 
\end{equation}
\end{Lemm}

 Note that $\mathbf{Q}_\eta(k)$ is a unit vector which depends on $k$ only through $k/|k|$,
and it can continuously be extended into the removable singularity at $k_3=-|k|$.
 Note also that $k\cdot \mathbf{Q}_\eta(k)=0$. 
 We remark that this is not in violation of the ``hairy ball theorem'' because $\mathbf{Q}_\eta(k)$ is \emph{complex}.

 With the help of Lemma \ref{lemm:Moses} we prove
\begin{Prop}\label{Prop:Gsplit}
 Suppose $\cG(V_h+\nabla\times w)<\infty$.
 Then, after at most a gauge transformation $w\mapsto w +\nabla\gamma$, we have
$w\in (\dot{W}_0^{\curl,1}(\Rset^3))^3+(\dot{W}_0^{\curl,2}(\Rset^3))^3$.
\end{Prop}

\noindent
\textit{Proof}: 

 For any $V=V_h+\nabla\times w$ with $\cG(V)<\infty$ the subset $K_C\subset\Rset^3$ on which $|V|\geq {C}>0\ a.e.$ 
has finite Lebesgue measure, for we have $\cG(V) \geq (\sqrt{1+{C}^2}-1)|K_C|$.

 Partitioning $\Rset^3 = K_C\cup \Rset^3\backslash K_C$ we thus estimate
\begin{equation}\tag{E17}\label{eq:GgeqNORMsum}
\cG(V)\geq \norm{|V|}_{L^1(K_C)}-|K_C|+
\tfrac{1}{1+\sqrt{1+C^2}}\norm{|V|}_{L^2(\Rset^3\setminus K_C)}^2.
\end{equation}

 Now let $C>0$ be small enough so that $\{s_l\}_{l=1}^N\subset\subset K_C$.
 Then, by the triangle inequality, (\ref{eq:GgeqNORMsum}) implies that, on the one hand,
\begin{equation}\tag{E18}
\|{|\nabla\times w|}\|_{L^1(K_C)} \leq \cG(V) + |K_C| + \norm{|V_h|}_{L^1(K_C)},
\end{equation}
and, on the other,
\begin{equation}\tag{E19}
\|{|\nabla\times w|}\|_{L^2(\Rset^3\setminus K_C)} 
\leq \sqrt{({1+\sqrt{1+C^2}})(\cG(V) + |K_C|)} + \norm{|V_h|}_{L^2(\Rset^3\setminus K_C)}.
\end{equation}
 Defining 
\begin{equation}\tag{E20}\label{eq:MOSESrepINVone}
f_{1;\eta}(k) := \int_{K_C} e^{-i2\pi k\cdot s}\mathbf{Q}^*_\eta(k)\cdot \nabla\times w(s) \dd^3s,
\end{equation}
\begin{equation}\tag{E21}\label{eq:MOSESrepINVtwo}
f_{2;\eta}(k) := \int_{\Rset^3\setminus K_C} e^{-i2\pi k\cdot s}\mathbf{Q}^*_\eta(k)\cdot \nabla\times w(s) \dd^3s,
\end{equation}
and $f_\eta(k) := f_{1;\eta}(k) + f_{2;\eta}(k)$, we obtain a decomposition of $\nabla\times w$ into
$\nabla\times w_1 + \nabla\times w_2$ with 
$|\nabla\times w_1|\in L^1(\Rset^3)$ and $|\nabla\times w_2|\in L^2(\Rset^3)$.
 But then (see \cite{Moses}),
\begin{equation}\tag{E22}\label{eq:MOSESrepSPLIT}
w_l(s): =
{\textstyle\sum\limits_{\eta\in\{\pm1\}}}\eta\int_{\Rset^3} e^{i2\pi k\cdot s}\mathbf{Q}_\eta(k) f_{l;\eta}(k) \tfrac{1}{|k|}\dd^3k
\in \dot{W}^{\curl,l}_0(\Rset^3)^3, \quad l\in\{1,2\}.
\end{equation}
\hfill \QED

 Next we use that the splitting $w=w_1+w_2$ is not unique to show:
\begin{Lemm}\label{lem:weakLtwoCMPCTness}
 Let $\{w^{(n)}\}_{n\in\Nset}\subset(\dot{W}_0^{\curl,1}(\Rset^3))^3 + (\dot{W}_0^{\curl,2}(\Rset^3))^3$ 
be a minimizing sequence for $\cG(V_h+\nabla\times w)$.
 Then without loss of generality we may assume that
$\||\nabla\times w_2^{(n)}|\|_{L^2} \leq C_2$ for some convenient fixed $C_2>0$. 
\end{Lemm}

\noindent
\textit{Proof}: 

 Since $w\in(\dot{W}_0^{\curl,1}(\Rset^3))^3 + (\dot{W}_0^{\curl,2}(\Rset^3))^3$ we have
$f_\eta \in \widetilde{C}_0(\Rset^3) + L^2(\Rset^3)$, where $\widetilde{C}_0(\Rset^3)$, a subset of the
continuous functions which vanish at spatial $\infty$, is the image of $L^1(\Rset^3)$ under Fourier transform.
 
 Now, if $f = f_1+f_2$ with $f_1 \in \widetilde{C}_0(\Rset^3)$ and $f_2\in L^2(\Rset^3)$, then for $h\in S(\Rset^3)$ 
(Schwartz space) also $f = (f_1+h) + (f_2-h)$ with $f_1+h \in\widetilde{C}_0(\Rset^3)$ and $f_2-h\in L^2(\Rset^3)$.
 Since furthermore $S(\Rset^3)$ is dense in $L^2(\Rset^3)$ (and also in $C_0(\Rset^3)$, though that's not needed), 
if necessary after splitting $w=w_1+w_2$ corresponding to $f_\eta = f_{1,\eta} +f_{2,\eta}$, 
we can always find an $h$ to ``re-split'' $w=w'_1+w'_2$ with $f_\eta = (f_{1,\eta}+h) + (f_{2,\eta}-h)$, 
such that $\||\nabla\times w'_2|\|_{L^2} \leq C_2$ for any fixed $C_2>0$. 
 This establishes the Lemma. \hfill \QED
\medskip

 Now consider any minimizing sequence $\{V^{(n)} = V_h +\nabla\times w^{(n)}\}_{n\in\Nset}$ 
of $G(V)$ with $w^{(n)}\in (\dot{W}_0^{\curl,1}(\Rset^3))^3+(\dot{W}_0^{\curl,2}(\Rset^3))^3$
satisfying $\||\nabla\times w_2|\|_{L^2} \leq C_2$.
 The $(\dot{W}_0^{\curl,2}(\Rset^3))^3$ norm bound on $w_2^{(n)}$ implies weak compactness, so from any minimizing sequence $\{V^{(n)}
= V_h +\nabla\times (w^{(n)}_1 + w^{(n)}_2)\}_{n\in\Nset}^{}$ we can extract a subsequence $\{V^{(n_j)}\}_{j\in\Nset}^{}$
such that $(\dot{W}_0^{\curl,2}(\Rset^3))^3$-weakly we have that $w^{(n_k)}_2\rightharpoonup w_2 \in (\dot{W}_0^{\curl,2}(\Rset^3))^3$ as 
$k\to\infty$; moreover, $\norm{\abs{\nabla\times w_2}}_{L^2}\leq {C}_2$.
 This reduces the problem to
proving weak compactness w.r.t. $(\dot{W}_0^{\curl,1}(\Rset^3))^3$ of the sequence $\{w^{(n)}_1\}_{n_k\in\Nset}^{}$ 
in a minimizing sequence $\{V^{(n)}\}_{n\in\Nset}^{}$ given by $V^{(n)} = V_h + \nabla\times (w^{(n)}_1 + w_2)$,
with $w_2$ denoting a weak limit point of the sequence $\{w^{(n)}_2\}_{n_k\in\Nset}^{}$ suitably chosen
in the original minimizing  sequence $\{V^{(n)}\}_{n\in\Nset}^{}$.
 The fact that $\{V^{(n)} = V_h + \nabla\times (w^{(n)}_1 + w_2)\}_{n_k\in\Nset}^{}$ is a minimizing sequence follows from
(\ref{eq:cGvminuscGVh}) with $V_1 = V_h + \nabla\times (w^{(n)}_1 + w_2^{(n)})$ and $V_2 = \nabla\times (w_2 - w_2^{(n)})$, 
which yields the estimate $\abs{\cG(V_1+V_2)-\cG(V_1)}\leq C \||{\nabla\times (w_2^{(n)}-w_2)}|\|_{L^2}$ 
(cf. (\ref{eq:ABScGvminuscGVhBOUNDb})).
 
 We now prove weak compactness w.r.t. $(\dot{W}_0^{\curl,1}(\Rset^3))^3$ of the sequence $\{w^{(n)}_1\}_{n_k\in\Nset}^{}$.

 First of all, since $w\equiv 0$ is not a minimizer if $N>1$, we have $\inf_V\!\cG(V)~<~\cG(V_h)$.
 Thus, without loss of generality we have 
$\cG(V_h^{} + \nabla\times w_2 + \nabla\times w^{(n)}_1 ) <\cG(V_h)$
and $\nabla\times w_2 + \nabla\times w^{(n)}_1 \not\equiv 0$, for all $n\in\Nset$.

 Second, recalling that for any $V=V_h+\nabla\times w$ with $w\in (\dot{W}_0^{\curl,1}(\Rset^3))^3+(\dot{W}_0^{\curl,2}(\Rset^3))^3$
the subset $K_C\subset\Rset^3$ on which $|V|\geq {C}>0\ a.e.$ has finite Lebesgue measure, since 
$\cG(V) \geq (\sqrt{1+{C}^2}-1)|K_C|$, without loss of generality we now choose $C=\sqrt{3}$ 
and thus may assume that when $N>1$, then for a minimizing sequence $\{V^{(n)}\}_{n\in\Nset}$ 
the size of the domains $K^{(n)}_{\surd{3}}$ on which $|V^{(n)}|\geq \surd{3}\ a.e.$ is bounded 
by $|K^{(n)}_{\surd{3}}| < \cG(V_h)$, uniformly in $n$.

 We also define $\Upsilon^{(n)} := K_{\surd{3}}^{(n)}\cup \overline{B_R}$, where
$B_R\supset\{s_k\}_{l=1}^N$ is an open ball and $\overline{B_R}$ its closure; then dist$(\partial B_R,\{s_l\}_{l=1}^N)>0$.
 Note that $|\Upsilon^{(n)}|\leq |B_R| +|K_{\surd{3}}^{(n)}|< |B_R| +  \cG(V_h)$ uniformly in $n$.

 Third, we establish the analog of the uniform upper norm bounds obtained in the proof of Proposition \ref{Prop:Gsplit} 
for the partitioning $\Rset^3 = \Upsilon^{(n)} \cup \Rset^3\backslash\Upsilon^{(n)}$.
 We estimate
\begin{equation}\tag{E23}\label{eq:GgeqNORMsumN}\hspace{-10pt}
\cG(V^{(n)})\geq \norm{|V^{(n)}|}_{L^1(\Upsilon^{(n)})}-|\Upsilon^{(n)}|+
\tfrac{1}{3}\norm{|V^{(n)}|}_{L^2(\Rset^3\setminus \Upsilon^{(n)})}^2
\end{equation}
and recall that $\cG(V_h^{}) > \cG(V^{(n)})$ when $N>1$.

 On the one hand (\ref{eq:GgeqNORMsumN}) implies
$\norm{|V^{(n)}|}_{L^1(\Upsilon^{(n)})} < \cG(V_h) + |\Upsilon^{(n)}|$.
 But then, since 
$\norm{|V^{(n)}|}_{L^1(\Upsilon^{(n)})}\geq \norm{|V^{(n)}|}_{L^1(B_R)}
\geq\|{|\nabla\times w^{(n)}|}\|_{L^1(B_R)}- \norm{|V_h|}_{L^1(B_R)}$, 
we obtain
$\|{|\nabla\times w^{(n)}|}\|_{L^1(B_R)} < \cG(V_h) + |\Upsilon^{(n)}| + \norm{|V_h|}_{L^1(B_R)}$; 
and since the triangle inequality, followed by a radially-decreasing-rearrangement inequality, gives us
$\norm{|V_h|}_{L^1(B_R)}\leq (4\pi)^{2/3}3^{1/3}(|B_R|)^{1/3}\sum_n|a_n|$, 
we conclude that (when $N>1$) 
$\| |\nabla\times w^{(n)}|\|_{L^1(B_R)}$ is bounded above uniformly in $n$.
 Also, since $\| |\nabla\times w_2|\|_{L^1(B_R)}<\infty$ independently of $n$,
for $w^{(n)}=w^{(n)}_1+w_2$ the triangle inequality now implies that 
$\|{|\nabla\times w^{(n)}_1|}\|_{L^1(B_R)} < \cG(V_h) + |\Upsilon^{(n)}| + \norm{|V_h|}_{L^1(\Upsilon^{(n)})}
 + \| |\nabla\times w_2|\|_{L^1(B_R)}\leq \overline{C}_1(R)$ uniformly in $n$.

 On the other hand, (\ref{eq:GgeqNORMsumN}) implies 
$\norm{|V^{(n)}|}_{L^2(\Rset^3\setminus \Upsilon^{(n)})}^2 < 3\big( \cG(V_h) + |\Upsilon^{(n)}|\big)$, 
which in concert with
$\norm{|V^{(n)}|}_{L^2(\Rset^3\setminus \Upsilon^{(n)})}
\geq \|{|\nabla\times w^{(n)}|}\|_{L^2(\Rset^3\setminus \Upsilon^{(n)})} - \norm{V_h}_{L^2(\Rset^3\setminus \Upsilon^{(n)})}$
yields the upper bound
$ \|{|\nabla\times w^{(n)}|}\|_{L^2(\Rset^3\setminus \Upsilon^{(n)})} 
\leq \sqrt{3\big( \cG(V_h) + |\Upsilon^{(n)}|\big)} + \norm{V_h}_{L^2(\Rset^3\setminus \Upsilon^{(n)})}$.
 Since $\norm{V_h}_{L^2(\Rset^3\setminus \Upsilon^{(n)})}\leq 4\pi \sum_k|a_k|/{\mbox{\textrm{dist}}(\partial B_R,\{s_j\}_{j=1}^N)}$,
and since $ \|{|\nabla\times w^{(n)}_1|}\|_{L^2(\Rset^3\setminus \Upsilon^{(n)})} 
\leq  \|{|\nabla\times w^{(n)}|}\|_{L^2(\Rset^3\setminus \Upsilon^{(n)})} +
 \|{|\nabla\times w_2|}\|_{L^2(\Rset^3\setminus \Upsilon^{(n)})}$
by the triangle inequality, and since $ \|{|\nabla\times w_2|}\|_{L^2(\Rset^3\setminus \Upsilon^{(n)})}\leq C_2$ by
Lemma \ref{lem:weakLtwoCMPCTness}, we conclude that for $N>1$, we have
$\| |\nabla\times w^{(n)}_1|\|_{L^2(\Rset^3\setminus \Upsilon^{(n)})}\leq \overline{C}_2$ uniformly in $n$.

 Fourth, by the $L^1(B_R)$ bounds a minimizing sequence $\{\big|{\nabla\times w^{(n)}_1}\big|\}_{n\in\Nset}$ 
has a weak$^*$ convergent subsequence on every $\overline{B_R}$; by the Lebesgue decomposition theorem
the limit is the sum of an $L^1(B_R)$ function $f$ and a measure $\mu$ which is singular w.r.t. Lebesgue measure,
with $\|f\|_{L^1(B_R)} + \mu(\overline{B_R}) \leq \overline{C}(R)$. 
 Now suppose that $\mu(\overline{B_R}) >0$; note that supp$(\mu)$ has Lebesgue measure zero.
 Then for any open neighborhood ${\mathcal{N}}_\eps$ of supp$(\mu)$ of size $\epsilon$ we have
\begin{equation}\tag{E24}\hspace{-10pt}
\int_{{\mathcal{N}}_\eps\cap B_R}\hspace{-5pt} \abs{V^{(n)}(s)}\dd^3s \geq 
\int_{{\mathcal{N}}_\eps\cap B_R}\!\! \Bigl(\sqrt{1 + \abs{V^{(n)}(s)}^2} -1\Bigr)\dd^3s \geq
\int_{{\mathcal{N}}_\eps\cap B_R}\hspace{-5pt} \abs{V^{(n)}(s)}\dd^3s -\eps.
\end{equation}
 Taking the limit $n_k\to\infty$ (along the convergent subsequence) and then $\eps\to 0$ reveals that
the singular part makes an additive contribution $\mu(B_R)>0$ to $G(V)$. 
 Thus by subtracting the part of the $\nabla\times w^{(n)}_1$ subsequence which converges in absolute
value to $\mu$ we can lower the value of $G(V)$; hence,
$\{{\nabla\times w^{(n)}_1}\}_{n\in\Nset}$ was not a minimizing sequence --- in contradiction to the
hypothesis that it was.
 Therefore, after extracting a subsequence, we can assume that a minimizing sequence
$\{{\nabla\times w^{(n)}_1}\}_{n\in\Nset}$ converges weakly in $L^1(B_R)$ to some $\nabla\times w_1$,
for each $B_R$ as stipulated. 
 This means that $\{\big|{\nabla\times w^{(n)}_1}\big|\}_{n\in\Nset}$ converges weakly in $\Lloc^1(\Rset^3)$ 
to some $\abs{\nabla\times w_1}$, and so 
$\{{\nabla\times w^{(n)}_1}\}_{n\in\Nset}$ converges weakly in $\dWloc^{\curl,1}(\Rset^3)^3$ 
to some ${\nabla\times w_1}$.

 Summarizing so far: any minimizing sequence $\{w^{(n)}\}_{n\in\Nset}^{}$ 
of $\cG(V_h+\nabla\times w)$ with $w^{(n)}\in (\dot{W}_0^{\curl,1}(\Rset^3))^3 + (\dot{W}_0^{\curl,2}(\Rset^3))^3$
has a locally weakly convergent subsequence with limit 
$w_* \in (\dWloc^{\curl,1}(\Rset^3))^3 + (\dot{W}_0^{\curl,2}(\Rset^3))^3$.
 
  Fifth, by another variation of the reasoning in our proof of Proposition \ref{Prop:Gsplit} 
we show that $w_*\in (\dot{W}_0^{\curl,1}(\Rset^3))^3 + (\dot{W}_0^{\curl,2}(\Rset^3))^3$.
  Namely, the subset $K_{\surd{3}}^*\subset\Rset^3$ on which $|V_h+\nabla\times w_*|\geq \surd{3}\ a.e.$ 
has finite Lebesgue measure, $|K^*_{\surd{3}}| < \cG(V_h)$.
 We also define $\Upsilon^* := K_{\surd{3}}^*\cup \overline{B_R}$ and
note that $|\Upsilon^*|\leq |B_R| +|K_{\surd{3}}^*|< |B_R| +  \cG(V_h)$.
 Essentially verbatim to our estimation of the sequence norms we have that 
$\|{|\nabla\times w_1|}\|_{L^1(\Upsilon^*)} < \cG(V_h) + |\Upsilon^*| + \norm{|V_h|}_{L^1(\Upsilon^*)}
 + \| |\nabla\times w_2|\|_{L^1(\Upsilon^*)}$.
 By H\"older's inequality, 
$\| |\nabla\times w_2|\|_{L^1(\Upsilon^*)}\leq |\Upsilon^*|^{1/2} \||\nabla\times w_2|\|_{L^2(\Upsilon^*)}^{1/2}$,
and we have $\||\nabla\times w_2|\|_{L^2(\Upsilon^*)}\leq \||\nabla\times w_2|\|_{L^2(\Rset^3)}$.
 And so $\|{|\nabla\times w_1|}\|_{L^1(\Upsilon^*)} \leq C_1$. 
 
 On the other hand, we also know that
$\| |\nabla\times w^{(n)}_1|\|_{L^2(\Rset^3\setminus \Upsilon^{(n)})}\leq \overline{C}_2$ uniformly in $n$, and
so 
$\| |\nabla\times w_1|\|_{L^2(\Rset^3\setminus \Upsilon^*)}\leq \overline{C}_2$.
 Now suppose that $\| |\nabla\times w_1|\|_{L^1(\Rset^3\setminus \Upsilon^*)}=\infty$.
 But then, since $|\nabla\times w_1| \in (L^2\cap L^\infty)(\Rset^3\setminus \Upsilon^*)$, 
we can subtract the offending part from $w_1$ and add it to $w_2$, denoting the new decomposition 
by $\tilde{w}_1+\tilde{w}_2$. 
 Indeed, by the chiral Fourier representation we can find a $\nabla\times\tilde{w}$ with chiral Fourier
components $\tilde{g}_\eta(k)\in (L^1\cap L^2)(\Rset^3)$ such that
$\| |\nabla\times (w_1-\tilde{w})|\|_{L^1(\Rset^3\setminus \Upsilon^*)}<\infty$; 
 since $|\nabla\times \tilde{w}| \in (L^2\cap L^\infty)(\Rset^3)$, it follows that
$|\nabla\times \tilde{w}| \in L^1(\Upsilon^*)$, and so 
$\| |\nabla\times (w_1-\tilde{w})|\|_{L^1(\Rset^3)}<\infty$. 
 Also we obviously have $|\nabla\times(w_2-\tilde{w})|\in L^2(\Rset^3)$. 
 Thus setting $\tilde{w}_1: = w_1 -\tilde{w}$ and $\tilde{w}_2 := w_2 +\tilde{w}$ we have 
$w_1+w_2 = \tilde{w}_1 +\tilde{w}$ with 
$\tilde{w}_1 \in \dot{W}_0^{\curl,1}(\Rset^3))^3$ and $\tilde{w}_2 \in \dot{W}_0^{\curl,2}(\Rset^3))^3$.
 Thus $w_* \in \dot{W}_0^{\curl,1}(\Rset^3))^3 + (\dot{W}_0^{\curl,2}(\Rset^3))^3$, as claimed.

 Lastly, by weak lower semi-continuity (Fatou's lemma),
\beq\tag{E25}\label{eq:Udef}
U(s): = V_h(s) +\nabla\times w_*(s)
\eeq
is a minimizer of $\cG(V)$.
 This completes the proof of Theorem \ref{theo:minEXISTS}. \hfill \QED

  It now follows in the usual way that $U$ given in (\ref{eq:Udef}) is a critical point of $\cG(V)$, 
satisfying the Euler--Lagrange equation
\beq\tag{E26}
\nabla\times \frac{V_h +\nabla\times w_*}{\sqrt{1 + \abs{V_h +\nabla\times w_*}{}^2_{}}} =0.
\eeq
      Thus, 
and applying the Poincar\'e lemma, we find that \emph{locally} in simply connected domains
\beq\tag{E27}\label{eq:EUdef}
E_U(s) = \frac{U(s)}{\sqrt{1 + \abs{U(s)}{}^2_{}}} \equiv -\nabla v_*(s)
\eeq
is a gradient field.
 Furthermore, as shown in \eqref{eq:divVasperEsubV} for any $E_V$, we have that 
\begin{equation}\tag{E28}\label{eq:ELeqn}
 - \nabla\cdot  \frac{\nabla v_*(s)}{\sqrt{1 - \abs{\nabla v_*(s)}{}^2_{}}}
= 
4\pi\sum_{n=1}^N a_n \delta_{s_n}
\end{equation}
in the sense of distributions; thus, $v_*$ exists globally $a.e.$ and
satisfies $\cF^{(1)}[v_*](\psi)=0$ for all $\psi\in C^\infty_c(\Rset^3)$.

 Finally, we show that $v_*=v_\infty$.
 Namely, we have
\beq\tag{E29}\label{eq:cFvstarIScFvmin}
	\cF(v_*)
 \geq 
	\cF(v_\infty)
 \geq 
-\cG(U) = \cF(v_*).
\eeq
 The first inequality expresses the fact that $v_\infty$ is a minimizer of $\cF(v)$; the second inequality
follows from the fact that $U$ is just a special $V$, and that $ \cF(v_\infty) \geq -\cG(V)$ for
all $V=V_h+\nabla\times w$, see \eqref{eq:cFinfABOVEnegcG}; lastly, the equality in \eqref{eq:cFvstarIScFvmin}
follows from the fact that 
$U = -\nabla v_*(s)\big/ \sqrt{1 - \abs{\nabla v_*(s)}{}^{2}_{}}$ satisfies \eqref{eq:ELeqn}, equivalently
$v_*$ satisfies $\cF^{(1)}[v_*](\psi)=0$ for all $\psi\in C^\infty_c(\Rset^3)$
 --- more explicitly, 
inverting the above stated algebraic relation between $U$ and $\nabla v_*$, we can rewrite $-\cG(U)$ as follows
(cf. (23)--(26) in [1]):
\beq\tag{E30}\label{eq:vstarISvmin}
 \displaystyle\int_{\Rset^3} \Bigl(1- \sqrt{1 + \abs{U(s)}{}^2}\Bigr)\dd^3s 
= \displaystyle\int_{\Rset^3} \Bigl(1- \sqrt{1 - \abs{\nabla v_*(s)}{}^2}
		+ U(s)\cdot  \nabla v_*(s)\Bigr)\dd^3s ,
\eeq
and an integration by parts on the last term in the second integral, and using that $\nabla\cdot U = 
4\pi\sum_{n=1}^N a_n \delta_{s_n}$ weakly, now yields $-\cG(U) = \cF(v_*)$.
 Thus, $\cF(v_*) = \cF(v_\infty)$, and by the uniqueness of the minimizer of $\cF(v)$, we have $v_*=v_\infty$.

 The proof that $\cF^{(1)}[v_\infty](\psi)=0$ for all $\psi\in C^\infty_c(\Rset^3)$ is complete.
\hfill \QED

\vfill
\noindent
\textbf{ACKNOWLEDGEMENT:} 
	My sincere thanks go to Denis Bonheure for raising his criticism of [1].
        I am also very grateful to the referees for their critical comments which helped me to 
improve this erratum. 
        I also thank Sagun Chanillo, Markus Kunze, and Shadi Tahvildar-Zadeh for their helpful comments.

\newpage

\section*{Appendix}
\subsection*{A. Proof of \eqref{eq:ABScGvminuscGVhBOUNDc}}

 Pick a ball $B_R\supset\{s_n\}_{n=1}^N$. 
 Also let $\Omega_{1}:= \{\abs{\nabla\times w_1}> 1\}$ and $\Omega_{2}:= \{\abs{\nabla\times w_2}> 1\}$.
 Then 
$\infty>\norm{\abs{\nabla\times w_1}}_{L^1}^{} \geq \int_{\Omega_{1}}\abs{\nabla\times w_1}d^3s \geq \abs{\Omega_{1}}$,
and similarly $\abs{\Omega_{2}}<\infty$.
 Let $\Omega = B_R \cup \Omega_{1} \cup \Omega_{2}$; then $\abs{\Omega}<\infty$.
 Let $\chi_S$ be the characteristic function of the set $S$. 
 Then, by the triangle inequality, 
$\norm{\abs{E_{V}}}_{L^2}^{}\leq 
\norm{\abs{E_{V}}\chi^{}_{\Omega}}_{L^2}^{}
+ \norm{\abs{E_{V}}\chi_{\Rset^3\setminus\Omega}}_{L^2}^{}$.
 Using $\abs{E_{V}}\leq 1$ yields
$\norm{\abs{E_{V}}\chi^{}_{\Omega}}_{L^2}^{}\leq |\Omega|^{1/2}$, 
while using $\abs{E_{V}}\leq \abs{V}$ and again the triangle inequality yields
\begin{equation}\tag{A1}\label{eq:EVestimateB}
\hspace{-10pt}
\norm{\abs{E_{V}}}_{L^2(\Rset^3\setminus\Omega)}^{}
\leq 
\norm{\abs{V_h}}_{L^2(\Rset^3\backslash \Omega)}+
\norm{\abs{\nabla\times w_1}}_{L^2(\Rset^3\backslash \Omega)}+
 \norm{\abs{\nabla\times w_2}}_{L^2(\Rset^3\backslash \Omega)}.
\end{equation}
 Next, $\abs{\nabla\times w_1}\leq 1$ on $\Rset^3\backslash \Omega$ yields
$\norm{\abs{\nabla\times w_1}}_{L^2(\Rset^3\backslash \Omega)}\leq 
\norm{\abs{\nabla\times w_1}}^{1/2}_{L^1(\Rset^3\backslash \Omega)}\leq 
\norm{\abs{\nabla\times w_1}}^{1/2}_{L^1}<\infty$. 
 Together with  $\norm{\abs{\nabla\times w_2}}_{L^2(\Rset^3\backslash \Omega)}
\leq  \norm{\abs{\nabla\times w_2}}_{L^2}<\infty$ 
we obtain \eqref{eq:ABScGvminuscGVhBOUNDc}.\hfill QED

%%%%%%%%%%%%%%%%%%%%%%%%%%%%%%%%%%%%%%%%%%%%%%%%%%%%%%%%%%%%%%%%%%%%%
		%%%%%%%%   BIBLIOGRAPHY  %%%%%%%%%%%